\documentclass[11pt]{article}
\usepackage{axodraw}
\usepackage{epsfig}
\usepackage{amsfonts}
\usepackage{amsmath}
\usepackage{bbm}
\input pix.sty

\newcommand{\tinymsbar}{{\overline{\mbox{\tiny\rm{MS}}}}}
\newcommand{\Lambdamsbar}{{\Lambda_\tinymsbar}}
\newcommand{\mD}{m_\rmi{D}}

\newcommand{\Nf}{N_{\rm f}}
\newcommand{\Nc}{N_{\rm c}}

\newcommand{\rmO}{{\mathcal{O}}}

\def\lsi{\raise0.3ex\hbox{$<$\kern-0.75em\raise-1.1ex\hbox{$\sim$}}}
\def\gsi{\raise0.3ex\hbox{$>$\kern-0.75em\raise-1.1ex\hbox{$\sim$}}}
\newcommand{\lsim}{\mathop{\lsi}}

\newcommand{\fe}{\rmi{f}}
\newcommand{\bo}{\rmi{b}}
\newcommand{\zfe}{\rmi{0f}}

\newcommand{\nF}[1]{n_\rmi{F{#1}}}
\newcommand{\nB}{n_\rmi{B}}
\newcommand{\rmii}[1]{{\mbox{\tiny\rm{#1}}}}
\newcommand{\re}{\mathop{\mbox{Re}}}
\newcommand{\im}{\mathop{\mbox{Im}}}

\newcommand{\Tint}[1]{{\hbox{$\sum$}\!\!\!\!\!\!\!\int\,}_{\!\!\!\!\raise-0.9ex\hbox{$\scriptstyle{#1}$}}}

\newcommand{\RR}{{\mathbb{R}}}
\newcommand{\ZZ}{{\mathbb{Z}}}

\newcommand{\unit}{{\mathbbm{1}}} %{\ii}
\newcommand{\bi}{\begin{itemize}}
\newcommand{\ei}{\end{itemize}}
\newcommand{\hide}[1]{ }
\newcommand{\bsl}[1]{\,\slash\!\!\!\!{#1}\,}

\makeatletter \@addtoreset{equation}{section} \makeatother
\renewcommand{\theequation}{\arabic{section}.\arabic{equation}}
\makeatletter
\renewcommand\section{\@startsection {section}{1}{\z@}%
                                   {-5.5ex \@plus -1ex \@minus -.2ex}% bfr-skip
                                   {2.3ex \@plus.2ex}%
                                   {\normalfont\large\bfseries}}
\renewcommand\subsection{\@startsection{subsection}{2}{\z@}%
                                     {-3.25ex\@plus -1ex \@minus -.2ex}%
                                     {1.5ex \@plus .2ex}%
                                     {\normalfont\normalsize\bfseries}}
\renewcommand\thesection {\@arabic\c@section}
\renewcommand\thesubsection   {\thesection.\@arabic\c@subsection}
\renewcommand{\@seccntformat}[1]{%
\csname the#1\endcsname.\hspace{1.0em}}
\makeatother
\begin{document}

\begin{titlepage}
\begin{flushright}
BI-TP 2007/35\\
arXiv:0711.1743\\ \vspace*{1cm}
\end{flushright}
\begin{centering}
\vfill

{\Large{\bf
% \\[2mm]
Heavy quarkonium in any channel in resummed hot QCD
}} 

\vspace{0.8cm}

Y.~Burnier, %%\footnote{yburnier@physik.uni-bielefeld.de}	  
M.~Laine, %%\footnote{laine@physik.uni-bielefeld.de}
M.~Veps\"al\"ainen %%\footnote{mtvepsal@physik.uni-bielefeld.de}	  

\vspace{0.8cm}

%$^\rmi{a}$%
{\em
Faculty of Physics, University of Bielefeld, 
D-33501 Bielefeld, Germany\\}

\vspace*{0.8cm}

\mbox{\bf Abstract}
 
\end{centering}

\vspace*{0.3cm}
 
\noindent
We elaborate on the fact that quarkonium in hot QCD should not be
thought of as a stationary bound state in a temperature-dependent real
potential, but as a short-lived transient, with an exponentially
decaying wave function. The reason is the existence of an imaginary
part in the pertinent static potential, signalling the
``disappearance'', due to inelastic scatterings with hard particles 
in the plasma, of the off-shell gluons that bind the quarks
together.  By solving the corresponding Schr\"odinger equation, 
we estimate numerically the near-threshold spectral functions in scalar,
pseudoscalar, vector and axial vector channels, as a function of the
temperature and of the heavy quark mass. In particular, we point out a
subtlety in the determination of the scalar channel spectral function
and, resolving it to the best of our understanding, suggest that at
least in the bottomonium case, a resonance peak {\em can} be observed
also in the scalar channel, even though it is strongly suppressed with
respect to the peak in the vector channel. Finally, we plot the
physical dilepton production rate, stressing that despite the eventual
disappearance of the resonance peak from the corresponding spectral
function, the quarkonium contribution to the dilepton rate becomes
{\em more pronounced} with increasing temperature, because of the
yield from free heavy quarks.

\vfill

%% %\noindent
%% %PACS numbers: 
%% %11.10.Wx, %        Finite temperature field theory
%% %11.15.Bt, %        General properties of perturbation theory
%% %11.15.Ha, %        Lattice gauge theory
%% %12.38.Bx, %        Perturbative calculations in QCD
%% %\\
%% %Keywords:
 
\vspace*{1cm}
  
\noindent
December 2007

\vfill

\end{titlepage}

%%%%%%%%%%%%%%%%%%%%%%%%%%% SECTION %%%%%%%%%%%%%%%%%%%%%%%%%%%%%%%%%%%%%
%
\section{Introduction}

Assuming the existence of a thermalized medium, with 
a temperature $T$, and of heavy quarks, with a mass $M \gg T$, 
there is a finite probability, given by the Boltzmann factor
$\exp(-2 M/T)$, that an on-shell quark and antiquark 
are generated through thermal fluctuations. They could
then annihilate, creating an off-shell photon, which may escape 
from the thermal system, and subsequently decay into a dilepton pair 
(for instance, $e^-e^+$ or $\mu^-\mu^+$). The characteristics of
the energy distribution of these pairs offer an indirect probe on 
the strongly interacting dynamics taking place within the thermalized system. 
As a concrete application, properties of lepton pairs can be 
observed in heavy ion collision experiments (see, e.g., refs.~\cite{exp}), 
and may serve as an indication of whether a thermalized state
with a temperature above the deconfinement transition was 
momentarily reached during the evolution~\cite{ms}. 

Given that various properties of the quarkonium system can be 
understood in great detail at zero temperature~\cite{quarkonium}, 
it could be assumed that describing quantitatively the heavy
quark-antiquark system in a thermalized medium 
is a relatively simple task. After all, QCD is asymptotically free, 
so the effective coupling should decrease with the temperature, 
and ultimately confinement is lost as well. Somewhat 
surprisingly, this expectation appears to be overly optimistic. 
In fact, all standard 
approximation methods develop further systematic uncertainties at $T > 0$. 
For instance, direct lattice QCD reconstructions of the quarkonium
{\em spectral function}~\cite{lattold,latt1,latt2}, 
which is a quantity determining the dilepton production rate, 
develop the new problem that an analytic continuation 
is needed from data collected
on a short Euclidean time interval, to the observable defined
in Minkowskian spacetime. Another popular class of approaches, 
so-called potential
models~\cite{models,mp3}, suffers from the proliferation of many 
independent non-perturbative definitions 
of a ``static potential'' which could be measured 
on the lattice~\cite{jp,lattpot} and inserted into 
a Schr\"odinger equation. 
A recently introduced method, the 
determination of the corresponding observable in strongly coupled 
$\mathcal{N} = 4$ Super-Yang-Mills theory~\cite{ads}, 
also contains unknown systematic 
errors from the point of view of QCD, which cannot be reduced by increasing 
the temperature, because the QCD coupling soon becomes weak~\cite{gE2}. 

The method employed in this paper is resummed weak-coupling
perturbation theory. It again suffers from novel difficulties at 
finite temperatures: curing infrared divergences necessitates
carrying out complicated resummations~\cite{dr,htl}, and even 
though a weak-coupling expansion in the QCD coupling constant $g$
can subsequently be defined, it has a strange structure, with 
relative corrections suppressed only by odd powers of $g$~\cite{jk,chm}, 
by logarithms like $g^n\ln(1/g)$~\cite{tt,mdlog}, 
or by powers of $g$ multiplied by
non-perturbative coefficients~\cite{linde}--\cite{nspt_mass}.
%%\cite{linde,mD,nspt_mass}.
Moreover, even if a number of coefficients were known, 
the convergence of the series could be slow~\cite{bn} 
(see, however, refs.~\cite{conv,gE2}).

Given all these problems, a suitable practical approach
at the present date might be to compute the quarkonium spectral 
function and the dilepton production rate with many different methods, 
possessing complementary systematic errors, and to look for a consistent
pattern, which could then also represent the situation in QCD. 
It is in this spirit that the purpose of the present paper
is to pursue the side of resummed perturbative computations. 

The resummed perturbative approach to heavy quarkonium
in hot QCD was initiated in 
refs.~\cite{static}--\cite{imV}, of which the present paper
is a direct continuation. In particular, we 
expand and improve on the analysis of ref.~\cite{og2}. 
We consider, first of all, the same observable as in ref.~\cite{og2} 
(quarkonium spectral function in the vector channel), but discuss more 
extensively the dependence of the result on the temperature and on 
the heavy quark mass. Second, we carry out a new analysis for the spectral 
function in the scalar channel. This turns out to require more advanced 
numerical techniques than those used in ref.~\cite{og2}. We also relate
the pseudoscalar and axial vector spectral functions to the vector 
and scalar spectral functions. 
Finally, we elaborate on the physics implications of the results
in more detail than before, 
both conceptually, i.e. with regard to the picture they 
suggest for the quarkonium system in a deconfined environment, and
from the practical point of view, i.e. with regard to the dilepton
production rate.

%%%%%%%%%%%%%%%%%%%%%%%%% SECTION %%%%%%%%%%%%%%%%%%%%%%%%%%%%%%%%%%%%%
%
\section{General framework}

We start by specifying somewhat more quantitatively 
the main ideas and equations of the resummed perturbative 
approach. A detailed derivation follows 
in \ses\ref{se:Schr}, \ref{se:rho}, while a reader only 
interested in the numerical results could skip directly
to \se\ref{se:num} after the present section.

Let $\hat\psi$ be a generic heavy quark field operator 
in the Heisenberg picture. 
The basic correlation function we consider is of the form
\be
  C_{>}^V(t;\vec{r,r'}) 
 \equiv % \frac{4}{9} e^2 \!
 \int \! {\rm d}^3 \vec{x}\,
 \Bigl\langle
  \hat{\bar\psi}\,\Bigl(t,\vec{x}+\frac{\vec{r}}{2}\Bigr)
  \gamma^\mu
  \, W 
  %% \Bigl[\Bigl(t,\vec{x}+\frac{\vec{r}}{2}\Bigr);
  %%                       \Bigl(t,\vec{x}-\frac{\vec{r}}{2}\Bigr)\Bigr]
  \, 
  \hat \psi\Bigl(t,\vec{x}-\frac{\vec{r}}{2}\Bigr) \;\; 
  \hat{\bar\psi}\,\Bigl(0,-\frac{\vec{r}'}{2}\Bigr)
  \gamma_\mu
  \, W' \,
  \hat{\psi}\Bigl(0,+ \frac{\vec{r}'}{2}\Bigr)
 \Bigr\rangle
 \;, \la{poinsplit}
\ee
where $W$, $W'$ are Wilson lines connecting
the adjacent operators,
inserted in order to keep the Green's function gauge-invariant;
the metric is $\eta_{\mu\nu} = \mathop{\mbox{diag}}$($+$$-$$-$$-$); 
and the expectation value refers to 
$\langle...\rangle\equiv \mathcal{Z}^{-1} \tr [\exp(-\hat H/T)(...)]$, 
where $\mathcal{Z}$ is the partition function, 
$\hat H$ is the QCD Hamiltonian, and $T$ is the temperature.
The superscript in $C^V_{>}$ refers to the vector channel; 
the subscript refers to the time-ordering in \eq\nr{poinsplit}.
We also consider scalar, pseudoscalar and axial vector correlators
below; their precise definitions are given 
in \se\ref{se:Schr}.\footnote{%
 In ref.~\cite{static}, we set $\vec{r}' = \vec{0}$, 
 and denoted the correlator by 
 $
  \check C_{>}(t,\vec{r}) \equiv 
  C_{>}^V(t;\vec{r,0}) 
 $.
 However, with certain channels, 
 it will be advantageous to keep $\vec{r}'\neq\vec{0}$, 
 because then the singularities from the static potential at 
 $\vec{r} = \vec{0}$, and from the initial condition 
 of the Schr\"odinger equation at $\vec{r} = \vec{r}'$,
 do not overlap.
 } 

The significance of the Green's function in \eq\nr{poinsplit} is
that if we take the limit $\vec{r,r'}\to\vec{0}$, and subsequently 
Fourier transform with respect to the time $t$, then we obtain
a function which is trivially related to the heavy quarkonium 
spectral function, $\rho^V(\omega)$, in the vector channel:
\be
 \rho^V(\omega) = \fr12 \Bigl( 1 - e^{-\frac{\omega}{T }}\Bigr)
 \int_{-\infty}^{\infty} \! {\rm d} t \, e^{i \omega t}
 \; C_{>}(t;\vec{0,0})
 \;. \la{rhoV}
\ee
This quantity is physically important, given that 
the production rate of $\mu^-\mu^+$ pairs (with a vanishing total
spatial momentum $\vec{0} = \vec{q}_{\mu^-} + \vec{q}_{\mu^+}$ 
and a non-vanishing total energy $\omega = E_{\mu^-} + E_{\mu^+}$) from
a system at a temperature $T$,  
is directly proportional to $\rho^V(\omega)$~\cite{dilepton}:
\be
 \frac{{\rm d} N_{\mu^-\mu^+}}{{\rm d}^4x\,{\rm d}^4 Q} = 
% \frac{{\rm d} \Gamma_{\mu^+\mu^-}}{{\rm d}^4 Q} = 
 \frac{2 c^2 e^4}{3 (2\pi)^5 \omega^2} % \theta(Q^2 - 4 m_\mu^2) 
 \biggl( 1 + \frac{2 m_\mu^2}{\omega^2}
 \biggr)
 \biggl(
 1 - \frac{4 m_\mu^2}{\omega^2} 
 \biggr)^\fr12 \nB(\omega) \Bigl[- \rho^V (\omega) \Bigr]
 \;, 
 \la{dilepton}
\ee
where we assumed $\omega \ge 2 m_\mu$; 
$e$ is the electromagnetic coupling;
$c\in(\fr23,-\fr13)$ is the electric 
charge of the heavy quark; and 
$\nB(\omega) \equiv 1/[\exp({\omega}/{T}) - 1]$
is the Bose distribution function. 
 
Now, a systematic perturbative determination of the Green's function 
in \eq\nr{poinsplit}, and of the corresponding spectral function in 
\eq\nr{rhoV},  is quite difficult for energies $\omega$
close to the quark-antiquark threshold, $\omega\sim 2 M$. The reason is 
that in this regime infinitely many graphs, particularly so-called ladders, 
contribute at the same order. A further problem is that at finite
temperatures, the rungs of the ladders, containing gluons, need to 
be dressed by thermal corrections. 

A way to resum these infinitely many dressed contributions is not to compute 
the correlator of \eq\nr{poinsplit} directly, but rather to find  
a partial differential equation satisfied by this correlator, and
then to solve this equation numerically. The partial differential
equation in question is just the Schr\"odinger equation. Indeed, 
it is for the sake of being able to write a Schr\"odinger equation 
that we have introduced $\vec{r,r'}\neq \vec{0}$ in \eq\nr{poinsplit}.  
To be more precise, let us consider $C_{>}^V$ in the 
limit that the heavy quark mass $M$ is very large. 
Then, as we will see, $C_{>}^V$ obeys
\be
 \biggl\{ i \partial_t - \biggl[ 2 M 
 + V_{>}(t,r)
 - \frac{\nabla_\vec{r}^2}{M}
 + \rmO\biggl(\frac{1}{M^2} \biggr)
 \biggr] \biggr\}  C_{>}^V(t;\vec{r,r'}) = 0 
 \;,  \la{Seq}
\ee
with the initial condition
\be
 C_{>}^V(0;\vec{r,r'}) = - 6 \Nc\, \delta^{(3)}(\vec{r-r'})
 + \rmO\biggl(\frac{1}{M} \biggr)
 \;, \la{In0}
\ee
where $\Nc=3$.
The terms specified 
explicitly in \eqs\nr{Seq}, \nr{In0} result from a tree-level 
computation; in contrast, 
the potential denoted by $V_{>}(t,r)$ originates only at 1-loop order. 
It can be defined as the coefficient scaling as $\rmO(M^0)$, 
after acting on $C_{>}^V(t;\vec{r,r'})$ with the time 
derivative $i\partial_t$. The potential $V_{>}(t,r)$ depends, 
in general, on the temperature; we assume that $T$ is parametrically
low compared with the heavy quark mass,  
$T\sim (g^2 ... g) M$ (cf.\ \se\ref{se:power}). 

Now, it can be argued that in order to be parametrically consistent, 
the static potential in \eq\nr{Seq} should be evaluated in the 
limit $t\gg r$ (cf.\ \se\ref{se:power}). 
%% sec.~3.1 of ref.~\cite{og2}). 
Then it obtains a simple form: in dimensional regularization
(cf.\ Eqs.~(4.3), (4.4) of ref.~\cite{static}),
\ba
 \lim_{t\to \infty} V_{>}(t,r) 
 & = & 
 -\frac{g^2 C_F}{4\pi} \biggl[ 
 \mD + \frac{\exp(-\mD r)}{r}
 \biggr] - \frac{i g^2 T C_F}{4\pi} \, \phi(\mD r)  + \rmO(g^4)    
 \;, \la{expl}
\ea
where  $C_F\equiv (\Nc^2-1)/2\Nc$;  
$\mD$ is the Debye mass parameter; 
and the function
\be
 \phi(x) \equiv 
 2 \int_0^\infty \! \frac{{\rm d} z \, z}{(z^2 +1)^2}
 \biggl[
   1 - \frac{\sin(z x)}{zx} 
 \biggr]
\ee
is finite and strictly increasing, 
with the limiting values $\phi(0) = 0$, $\phi(\infty) = 1$.

The first term in \eq\nr{expl} corresponds to 
twice a thermal mass correction for the heavy quarks
(cf.\ the first term inside the square brackets in \eq\nr{Seq}). 
The second term is a standard $r$-dependent Debye-screened potential. 
The third term represents an imaginary part: its physics is that 
almost static (off-shell) gluons may disappear due to inelastic
scatterings with hard particles in the plasma. This is the phenomenon 
of Landau-damping, well-known in plasma physics. As a consequence
of the imaginary part, the solution of the Schr\"odinger equation
does not lead to a stationary wave function: rather, the bound state 
decays exponentially with time, representing a short-lived transient. 

In the following two sections, we discuss the origin of the formulae presented
here, and their practical evaluation, in some more detail. We also 
extend the discussion to the other channels. We return to the numerical 
solution of the Schr\"odinger equation in \se\ref{se:num}.

%%%%%%%%%%%%%%%%%%%%%%%%% SECTION %%%%%%%%%%%%%%%%%%%%%%%%%%%%%%%%%%%%%
%
\section{Schr\"odinger equation and initial conditions}
\la{se:Schr}

Our strategy for the derivation of the Schr\"odinger 
equation satisfied by the two-point correlation functions in various 
channels will be quite straightforward and ``modest'' here\footnote{
 A more systematic approach might follow by generalizing
 the framework of PNRQCD~\cite{pnrqcd} to finite $T$.
 }:  
we first compute the correlation functions in tree-level 
perturbation theory, and then expand in inverse powers of the 
heavy-quark mass. At this point we can identify the Schr\"odinger-equation
and the initial condition for its solution. Subsequently, 
radiative corrections are expected to multiplicatively correct
the terms that already appear at tree-level, and to add other terms
which are allowed by symmetries, even if they would not appear at
tree-level; the most important of these is the static potential. 
As long as there is a hierarchy between the different physical 
scales relevant for the problem (cf.\ \se\ref{se:power}), 
and we are only sensitive to 
perturbative scales, general principles suggest that the system 
should remain local in the presence of radiative corrections, and
that a truncation to a certain order is possible. 

%%%%%%%%%%%%%%%%%%%%%%%%% SUBSECTION %%%%%%%%%%%%%%%%%%%%%%%%%%%%%%%%%%
%
\subsection{Power counting}
\la{se:power}

Let us consider the parametric orders of magnitude of the various
terms in \eq\nr{Seq}, given the potential in \eq\nr{expl}. 
We recall, first of all, that the term $2 M$
plays no role, since it can always be eliminated through a trivial
phase factor (cf.\ \eq\nr{psi_def}). Around the quarkonium peak, the 
time derivative (or energy) is then of the order of the kinetic terms, i.e.\
$
 \partial_t \sim \partial_r^2 / M
$. 
If we, furthermore, equate kinetic energy with the Coulomb 
potential energy (assuming $\mD r \lsim 1$, cf.\ below), we are lead to 
\be
 \partial_r \sim \frac{1}{r} \sim g^2 M \;, 
 \quad 
 \partial_t \sim \frac{1}{t} \sim g^4 M
 \;.  \la{param_magn}
\ee
An essential question is now to decide how the temperature, $T$, 
is to be compared with these scales. Let us assume, first of all, that 
\be
 T \sim g^2 M \qquad (\mbox{case 1})
 \;. 
\ee
Then $\mD r \sim g T r \sim g \ll 1$, and Debye screening plays 
essentially no role yet: we may assume the bound state to exist. 
In this limit, 
\be
 \re V_{>} \sim \frac{g^2}{r} \sim g^4 M \; \gg \;
 \im V_{>} \sim g^2 T (\mD r)^2 \sim g^6 M
 \;, 
\ee
and the imaginary part can indeed be neglected. 

On the other hand, let us increase the temperature to 
\be
 T \sim g M \qquad (\mbox{case 2})
 \;. 
\ee
Then Debye screening plays an essential role,
$\mD r \sim g T r \sim 1$,  
and we may assume that the bound state has melted:
indeed, in this limit, 
\be
 \re V_{>} \sim \frac{g^2}{r} \sim g^4 M \; \ll \;
 \im V_{>} \sim g^2 T \sim g^3 M
 \;, 
\ee
so that the imaginary part of the potential, or the width of the state, 
dominates over the real part of the potential, or the binding energy.

To summarise, the interesting temperature 
range is $T \sim (g^2 ... g) M$. In principle, parametrically consistent 
analyses in the two limiting cases may require different methods. 
In practice, we would like to have phenomenological access to the whole
range; therefore, in the present paper we work (implicitly) in the situation 
where $\re V_{>} \sim \im V_{>}$, setting us somewhere in the middle
of the range. For further reference, let us point out that in this 
situation, $r \nabla A \sim r \mD A \lsim A$, where $A$ is some 
gauge field component: the variation of the infrared gauge fields is 
parametrically small on the length scales set by the bound state radius. 

%%%%%%%%%%%%%%%%%%%%%%%%% SUBSECTION %%%%%%%%%%%%%%%%%%%%%%%%%%%%%%%%%%
%
\subsection{Vector channel}
\la{ss:rhoV}

Denoting 
\be
 V^\mu(x;\vec{r}) \equiv 
 \bar\psi\Bigl(t,\vec{x}+\frac{\vec{r}}{2}\Bigr) 
 \gamma^\mu  \, W
 \psi\Bigl(t,\vec{x}-\frac{\vec{r}}{2}\Bigr)
 \;,  \la{Vmudef}
\ee
where $x\equiv (t,\vec{x})$, the vector channel 
correlator we consider is in general of the type
\be
 C_{>}^{V}(x;\vec{r},\vec{r}') = 
 \Bigl\langle
   V^\mu(x;\vec{r}) V_\mu(0;-\vec{r}')
 \Bigr\rangle
 \;. \la{CV}
\ee
For simplicity, we have left out hats from the fields
in \eq\nr{Vmudef}, 
as is appropriate once we go over to the path integral formulation
in Euclidean spacetime. 

%\newpage

Now, even though we will carry out the computation of \eq\nr{CV}
within QCD below, 
it will be useful to rewrite the operators considered in the language 
of NRQCD~\cite{nrqcd0} (for a review, see ref.~\cite{nrqcd}), 
because this allows to immediately see their
scaling with the heavy quark mass~$M$, and because this allows to 
relate various operators to each other in the large-$M$ limit.  
Following ref.~\cite{fwt}, 
we can start by carrying out a Foldy-Wouthuysen transformation, 
\be
 \psi \longrightarrow 
 \exp\biggl( \frac{i \gamma^j \overrightarrow{\!D}_{\!\!j} }{2 M} \biggr) \psi
 \;, \quad
 \bar\psi \longrightarrow  
 \bar\psi\exp\biggl( 
 - \frac{i \gamma^j \overleftarrow{\!D}_{\!\!j} }{2 M} \biggr) 
 \;, \la{fw}
\ee
where $\overrightarrow{\!D}_{\!\!j} 
\,\equiv\, \overrightarrow{\!\partial}_{\!\!j}\! - i g A_j$, 
$\overleftarrow{\!D}_{\!\!j} 
\,\equiv\, \overleftarrow{\!\partial}_{\!\!j}\! + i g A_j$, 
and we assume a summation over spatial 
indices, $j=1,2,3$. Afterwards, we go over to 
the non-relativistic two-component
notation by writing 
\be
 \psi \equiv  
  \left( 
  \begin{array}{c} 
    \theta \\ \phi
  \end{array}
 \right)
 \;, \quad
 \bar\psi \equiv
 ( \theta^\dagger \;, \; - \phi^\dagger ) 
 \;, \la{nr}
\ee
where we already assumed 
a representation for the Dirac matrices with
\be
 \gamma^0 \equiv
 \left( 
  \begin{array}{cc}
   \unit & 0 \\   
   0 & -\unit 
  \end{array}
 \right)
 \;, \quad
 \gamma^k \equiv
 \left( 
  \begin{array}{cc}
   0 & \sigma_k \\   
   -\sigma_k & 0 
  \end{array}
 \right)
 \;, \quad k = 1,2,3
 \;. \la{Dirac_NRQCD}
\ee
Here $\sigma_k$ are the Pauli matrices. 
Furthermore, it is useful to note that in NRQCD, the actions
for $\phi$ and $\theta$ are of first order in time derivatives; 
consequently, one of the degrees of freedom propagates strictly 
forward in time, the other strictly backward in time, and 
a non-zero mesonic correlator at $t\neq 0$ is only obtained from structures 
like $\langle \phi^\dagger(...)\theta \; \theta^\dagger(...)\phi \rangle$.

We now find that for $V^0$, 
the leading term with the desired structure is $\rmO(1/M)$
(this term is also a total derivative in the limit $\vec{r}\to\vec{0}$).
Therefore, the correlator $C_{>}^{V}$ is dominated
by the spatial components $V^k$. 
At $\rmO(M^0)$, these become
\be
 V^k(x;\vec{r}) = 
 \theta^\dagger\Bigl( t,\vec{x}+ \frac{\vec{r}}{2} \Bigr)
 \sigma_k \, W 
 \phi \Bigl( t,\vec{x} - \frac{\vec{r}}{2} \Bigr)
 + 
 \phi^\dagger\Bigl( t,\vec{x}+ \frac{\vec{r}}{2} \Bigr)
 \sigma_k \, W 
 \theta \Bigl( t,\vec{x} - \frac{\vec{r}}{2} \Bigr)
 \;. \la{Vk_nr}
\ee 
To the extent that interactions between the quark and antiquark 
are spin-independent (this is violated only at $\rmO(1/M)$),
the Pauli-matrices play a trivial role in the two-point correlator
made out of these operators,
yielding  eventually  $\tr[\sigma_k \sigma_l]$, if $V^k$ and $V^l$ are being 
correlated. 
% Therefore, in the language of NRQCD, it is conventional 
% to say that $V^k(x;\vec{r})$ represents an \emph{S-wave} state: 
% it contains no spatial derivatives, or a preferred direction.

% \subsubsection{Tree-level computation}
% \la{se:tree_CV}

We now proceed to compute the correlator in \eq\nr{CV} at tree-level.
We start in Euclidean spacetime\footnote{%
 Since both Euclidean and Minkowskian objects appear in this paper, 
 we try to distinguish between them by denoting the former with a tilde.
 In particular, $\tilde P = (\tilde p_\zfe,\mathbf{p})$ 
 denotes fermionic Euclidean four-momenta, 
 while $\tilde\gamma_\mu$ stand for Euclidean 
 Dirac matrices, satisfying $\{\tilde \gamma_\mu, \tilde \gamma_\nu \} = 
 2 \delta_{\mu\nu}$. Any further
 unspecified conventions can be found in ref.~\cite{static}.
 } 
and after a spatial Fourier transform
(for the moment we keep, for generality, 
the spatial momentum non-zero, $\vec{q}\neq \vec{0}$, 
unlike in \eq\nr{poinsplit}), whereby
\ba
 & & \hspace*{-1cm} C_E^{V}(\tau,\vec{q};\vec{r},\vec{r}') 
 \nn & \equiv & 
 \int \! {\rm d}^3 \vec{x}\, e^{-i \vec{q}\cdot \vec{x}}
 \Bigl\langle
  {\!\bar\psi}\,\Bigl(\tau,\vec{x} +\frac{\vec{r}}{2}
  \Bigr)
  \gamma^\mu
  \, % W_{\vec{r}}[(t,\vec{x}+\frac{\vec{r}}{2});
     %                    (t,\vec{x}-\frac{\vec{r}}{2})] \, 
  \psi\Bigl(\tau,\vec{x}-\frac{\vec{r}}{2}
  \Bigr) \;\; 
  {\!\bar\psi}\,\Bigl(0,-\frac{\vec{r}'}{2}\Bigr)
  \gamma_\mu
  {\psi}\Bigl(0,+\frac{\vec{r}'}{2}\Bigr)
 \Bigr\rangle
 \la{Spoinsplit} \\[2mm]
  &  = & \!\!\!
 - \Nc \int \! {\rm d}^3 \vec{x} \, e^{-i \vec{q}\cdot \vec{x}}
   \Tint{\tilde P_\fe, \tilde S_\fe}
   e^{i (\tilde p_\zfe - \tilde s_\zfe)\tau + i (\vec{s}-\vec{p})\cdot \vec{x}
                    + i (\vec{s}+\vec{p})\cdot \frac{\vec{r-r'}}{2}
     } \;
   \tr\!
   \biggl[
    \tilde \gamma_\mu \frac{-i \bsl{\tilde P} + M }{\tilde P^2 + M^2} 
    \tilde \gamma_\mu \frac{-i \bsl{\tilde S} + M }{\tilde S^2 + M^2} 
   \biggr]   
 \nn[2mm] & = & \!\!\!
 - 8 \Nc 
   \Tint{\tilde P_\fe, \tilde S_\fe}
   (2\pi)^3 \delta^{(3)}(\vec{s} - \vec{p}-\vec{q})
   e^{i (\tilde p_\zfe - \tilde s_\zfe)\tau 
      + i (2 \vec{p} + \vec{q}) \cdot \frac{\vec{r-r'}}{2}
   } \;
   \frac{\tilde p_\zfe \tilde s_\zfe + \vec{p}\cdot \vec{s} + 2 M^2 }
        {(\tilde P^2 + M^2)(\tilde S^2 + M^2)}
  \nn[2mm] & = & \!\!\!
 - 4 \Nc 
   T^2 \!\! \sum_{\tilde p_\zfe, \tilde s_\zfe} \int \! 
   \frac{{\rm d}^3 \vec{p}}{(2\pi)^3} 
   e^{i (\tilde p_\zfe - \tilde s_\zfe)\tau 
      + i (2 \vec{p} + \vec{q}) \cdot \frac{\vec{r-r'}}{2}
   } \;
   \frac{
    2 \tilde p_\zfe \tilde s_\zfe 
  + E_\vec{p}^2 + E_\vec{p+q}^2 + 2 M^2 - \vec{q}^2 
        }
        {(\tilde p_\zfe^2 + E_\vec{p}^2)
         (\tilde s_\zfe^2 + E_\vec{p+q}^2)}
%  \times \nn & & \times  
%  \Bigl( 
%  \Bigr)
   \;, \hspace*{1cm}
\ea
where $\tilde p_\zfe = 2\pi T(n+\fr12) - i \mu$, 
$n \in \ZZ$, denotes fermionic Matsubara frequencies
($\mu$ is the quark chemical potential), 
and we have introduced the notation
\be
 E_\vec{p} \equiv \sqrt{M^2 + \vec{p}^2}
 \;.
\ee 
The Matsubara sums can be carried out, by making use of 
\ba
 T\sum_{\tilde p_\zfe} 
 \frac{e^{\pm i \tilde p_\zfe \tau}}{\tilde p_\zfe^2 + E^2} 
 & = & 
 \frac{1}{2E} \Bigl[ n_\rmi{F}(E\pm\mu)
 e^{(\beta-\tau)E \pm \beta\mu} - n_\rmi{F}(E\mp\mu) e^{\tau E}
 \Bigr]  
 \;, \la{Msum1} \\
 T\sum_{\tilde p_\zfe} \frac{\pm i \tilde p_\zfe 
 e^{\pm i \tilde p_\zfe \tau}}{\tilde p_\zfe^2 + E^2} 
 & = & 
 -\frac{1}{2} \Bigl[ n_\rmi{F}(E\pm\mu)
 e^{(\beta-\tau)E \pm \beta\mu} + n_\rmi{F}(E\mp\mu) e^{\tau E}
 \Bigr]  
 \;, \la{Msum2}
\ea
valid for $0<\tau< \beta$.
This yields
\ba
 && \!\!\!\! C_E^V(\tau,\vec{q};\vec{r,r'}) = 
 - \Nc \int \! \frac{{\rm d}^3\vec{p}}{(2\pi)^3} e^{i (\vec{2 p + q })\cdot
   \frac{\vec{r-r'}}{2}}
 \times \nn && \times \biggl\{  
%%%
 \nF{}(E_\vec{p}+\mu) \nF{}(E_\vec{p+q}-\mu)
 e^{(\beta-\tau)(E_\vec{p}+E_\vec{p+q})}
 \biggl[ 
% 2 + \frac{-\vec{q}^2 + 2 M^2 + E_\vec{p}^2 + E_\vec{p+q}^2}
%          {E_\vec{p}E_\vec{p+q}} 
 \frac{-\vec{q}^2 + 2 M^2 + (E_\vec{p} + E_\vec{p+q})^2}
          {E_\vec{p}E_\vec{p+q}} 
 \biggr] 
 + \nn && +     
%%%
  \nF{}(E_\vec{p}+\mu) \nF{}(E_\vec{p+q}+\mu)
 e^{(\beta-\tau)E_\vec{p}+\tau E_\vec{p+q}+\beta\mu}
 \biggl[ 
% 2 - \frac{-\vec{q}^2 + 2 M^2 + E_\vec{p}^2 + E_\vec{p+q}^2}
%          {E_\vec{p}E_\vec{p+q}} 
 \frac{\vec{q}^2 - 2 M^2 - (E_\vec{p} - E_\vec{p+q})^2}
          {E_\vec{p}E_\vec{p+q}} 
 \biggr] 
 + \nn && +     
%%%
  \nF{}(E_\vec{p}-\mu) \nF{}(E_\vec{p+q}-\mu)
 e^{(\beta-\tau)E_\vec{p+q}+\tau E_\vec{p}-\beta\mu}
 \biggl[ 
% 2 - \frac{-\vec{q}^2 + 2 M^2 + E_\vec{p}^2 + E_\vec{p+q}^2}
%          {E_\vec{p}E_\vec{p+q}} 
 \frac{\vec{q}^2 - 2 M^2 - (E_\vec{p} - E_\vec{p+q})^2}
          {E_\vec{p}E_\vec{p+q}} 
 \biggr] 
 + \nn && +     
%%%
 \nF{}(E_\vec{p}-\mu) \nF{}(E_\vec{p+q}+\mu)
 e^{\tau(E_\vec{p}+E_\vec{p+q})}
 \biggl[ 
% 2 + \frac{-\vec{q}^2 + 2 M^2 + E_\vec{p}^2 + E_\vec{p+q}^2}
%          {E_\vec{p}E_\vec{p+q}} 
 \frac{-\vec{q}^2 + 2 M^2 + (E_\vec{p} + E_\vec{p+q})^2}
          {E_\vec{p}E_\vec{p+q}} 
 \biggr] 
 \biggr\}
 \;. \la{SqcdCE}
\ea

In order to simplify the expression somewhat,
we note that once we go over into the spectral function\footnote{%
  Take first a Fourier transform, 
  $ 
    \tilde C_E(\omega_\bo) = 
    \int_0^\beta \! {\rm d}\tau \, e^{i \omega_\bo \tau}
     C_E(\tau)
  $, 
  where $\omega_\bo$ is a bosonic Matsubara frequency;
  then carry out the analytic continuation
  $
   \rho(\omega)
    = \frac{1}{2i} [ 
    \tilde C_E(-i[\omega+i 0^+]) - 
    \tilde C_E(-i[\omega-i 0^+])
    ]
  $.
  A typical term in $C_E(\tau)$, of the form 
  $
    \exp(\Delta_1 \tau +
       \Delta_2 (\beta-\tau))
  $,
  becomes 
  $
     \rho(\omega) =  
    - \pi ( e^{\beta \Delta_1} - e^{\beta \Delta_2} ) 
       \delta(\omega + \Delta_1 - \Delta_2)
  $. \la{rhorec}
  },
and restrict to frequencies (energies) around the quark-antiquark 
threshold, $|\omega - 2M| \ll M$, then only the first of 
the structures in \eq\nr{SqcdCE} contributes. Second,  
close enough to the threshold, the $\delta$-function expressing 
energy-conservation, $\delta(\omega - E_\vec{p} - E_\vec{p+q})$, 
forces the loop momentum $\vec{p}$ to be small, $|\vec{p}| \ll M$. 
We also assume the external momentum to be small, $|\vec{q}| \ll M$.
Under these circumstances, we can expand 
\be
 E_\vec{p} \approx M + \frac{\vec{p}^2}{2 M}
 \;, \quad
 E_\vec{p+q} \approx M + \frac{|\vec{p+q}|^2}{2 M}
 \;,
\ee
and the relevant part of $C_E^{V}(\tau,\vec{q};\vec{r},\vec{r}')$ becomes
\ba
 && \hspace*{-1cm} C_E^{V}(\tau,\vec{q};\vec{r},\vec{r}')
 \nn & \simeq &
 - 6 \Nc \int \! \frac{{\rm d}^3\vec{p}}{(2\pi)^3} 
  e^{i (2 \vec{p} + \vec{q}) \cdot \frac{\vec{r-r'}}{2} %%}
% \times \nn && \times \biggl\{  
%%%
 %% \nF{}(E_\vec{p}+\mu) \nF{}(E_\vec{p+q}-\mu)
%% e^{
  -\tau \bigl[ 2 M + 
 \frac{2 \vec{p}^2 + 2 \vec{p}\cdot \vec{q} + \vec{q}^2}{2 M}
 + \rmO\bigl(\frac{1}{M^3}\bigr) \bigr]}
 \biggl[ 1 + 
 \rmO\biggl( \frac{1}{M^2} \biggr)
 % 3 - 
 % \frac{ \vec{p}^2 + \vec{p}\cdot \vec{q} + \vec{q}^2}{M^2}
 \biggr] 
 %% \biggr\}
 \;. \hspace*{0.5cm} \la{Ssimple}
\ea
Here we have also omitted effects of relative order $\exp(-[M\pm\mu]/T)$, 
by keeping only the leading terms in the exponentials. 
We note that after these simplifications, all dependence on the 
temperature and on the chemical potential 
has disappeared from the tree-level result.

% \subsubsection{Representation through a Schr\"odinger-equation}

The real-time object we are ultimately interested in, 
is the analytic continuation 
\be 
 C_{>}^{V}(t,\vec{q};\vec{r,r'}) = C_E^{V}(it,\vec{q};\vec{r,r'})
 \;. \la{Scont}
\ee
Noting from \eq\nr{Ssimple} that 
\be
 -i \nabla_\vec{r} \Leftrightarrow \vec{p} + \frac{\vec{q}}{2}
 \;, 
\ee
the dependence on $\vec{r}$ and $t$ in the exponential 
amounts to satisfying the Schr\"odinger equation 
\be
 \biggl\{ i \partial_t - \biggl[ 2 M + 
 \frac{\vec{q}^2}{4 M}
 - \frac{\nabla_\vec{r}^2}{M}
 + \rmO\biggl(\frac{1}{M^3} \biggr)
 \biggr] \biggr\} C_{>}^{V}(t,\vec{q};\vec{r,r'}) = 0 
 \;.
 \la{SSchr}
\ee
The initial condition for the solution is obtained by
setting $t=0$ in \eq\nr{Ssimple} (after use of \eq\nr{Scont}): 
we find
\ba
 C_{>}^{V}(0,\vec{q};\vec{r,r'}) & = &  - 6 \Nc 
 \int \! \frac{{\rm d}^3\vec{p}}{(2\pi)^3} 
 e^{i (2 \vec{p} + \vec{q}) \cdot \frac{\vec{r-r'}}{2} }
 + \rmO\biggl(\frac{1}{M^2} \biggr)
 \nn & = & 
 - 6 \Nc \,
 \delta^{(3)}(\vec{r-r'})
 + \rmO\biggl(\frac{1}{M^2} \biggr)
 \;.
 \la{Sinit0}
\ea
\eqs\nr{SSchr}, \nr{Sinit0} justify 
\eqs\nr{Seq}, \nr{In0} for the vector channel in the free limit. 

For future reference, let us also 
compute $\rho^V(\omega)$ explicitly (general expressions for 
free spectral functions can be found in refs.~\cite{free}). 
\eq\nr{Ssimple} (after $\tau\to it$) already
shows the solution of \eqs\nr{SSchr}, \nr{Sinit0}, 
and we can then directly remove the point-splitting, 
setting $\vec{r,r'} = \vec{0}$.  Shifting $\vec{p}\to \vec{p} - \vec{q}/2$; 
taking the steps in footnote~\ref{rhorec}; and ignoring exponentially
small terms and terms suppressed by $\rmO(1/M^2)$, we find
\ba
 \rho^V(\omega) & \approx & - {6\Nc\pi}
 \int \! \frac{{\rm d}^3\vec{p}}{(2\pi)^3}
 \delta\Bigl( \omega' - \frac{\vec{p}^2}{M}\Bigr)
 = % \nn & = & 
 -\frac{3 \Nc }{2 \pi}
 \, \theta(\omega') \,
       M^{\fr32} (\omega')^\fr12  
 \;, \la{tree_S}
\ea
where
\be
 \omega' \equiv \omega - \Bigl[ 2 M + \frac{\vec{q}^2}{4 M} \Bigr] 
 \;. \la{omegap}
\ee
In the following, we will often
for simplicity restrict to $\vec{q} = \vec{0}$
(like already in \eq\nr{Seq}), but we  can now
observe from \eqs\nr{tree_S}, \nr{omegap} that the main 
effect of a non-zero $\vec{q} \neq \vec{0}$ is simply 
to shift the threshold location $2M$ by the center-of-mass 
kinetic energy $\vec{q}^2/4 M$.

The analysis so far has been at tree-level. 
As argued in refs.~\cite{static,og2}, however, the essential
(temperature and $\omega$-dependent) 1-loop corrections 
can be taken into account simply by inserting the potential 
$V_{>}(\infty,r)$, given in \eq\nr{expl}, into \eq\nr{SSchr}.
There are of course also other loop corrections, related for
instance to the renormalization and definition of $M$ as a pole 
mass, and the overall normalization of the non-relativistic vector 
current in \eq\nr{Vk_nr}; these corrections are in fact known to high 
order at zero temperature~\cite{cs,current},\footnote{%
 To 1-loop order, 
 $
  M = m_{\tinymsbar}(m_{\tinymsbar}) 
  (1 + g^2 C_F/4\pi^2)
 $, 
 $
  V^k_\rmii{NRQCD}(x;\vec{0}) = 
  V^k_\rmii{QCD}(x;\vec{0}) 
  (1 + g^2 C_F/2\pi^2)
 $.
 } 
but are not essential at our current
resolution, so we mostly omit them here.

%%%%%%%%%%%%%%%%%%%%%%%%% SUBSECTION %%%%%%%%%%%%%%%%%%%%%%%%%%%%%%%%%%
%
\subsection{Scalar channel}
\la{ss:rhoS}

Denoting 
\be
 S(x;\vec{r}) \equiv 
 \bar\psi\Bigl(t,\vec{x}+\frac{\vec{r}}{2}\Bigr) 
 W
 \psi\Bigl(t,\vec{x}-\frac{\vec{r}}{2}\Bigr)
 \;, 
\ee
the scalar channel correlator we consider is of the type
\be
 C_{>}^{S}(x;\vec{r},\vec{r}') = 
 \Bigl\langle
   S (x;\vec{r}) S (0;-\vec{r}')
 \Bigr\rangle
 \;. \la{CS}
\ee
The correlator $C_{>}^{S}(x;\vec{0},\vec{0})$ is not directly
physical\footnote{%
 It may be noted, for instance, 
 that the scalar density requires renormalization,
 unlike the vector current.
 }, 
but it does have the appropriate quantum numbers
to give a contribution to the three-particle
production rate $q\bar q\to \mu^-\mu^+\gamma$, 
i.e.\ a lepton--antilepton pair 
together with an on-shell photon. Moreover, it is frequently measured 
on the lattice, which will be our most direct reference point. 
We will ignore the issue of overall (re)normalization in the following, 
and concentrate on the shape of the spectral function (meaning
its $\omega$-dependence
in frequency space, or its $t$-dependence in coordinate space).

It is again helpful to write $S (x;\vec{r})$ with the NRQCD notation. 
The steps in \eqs\nr{fw}, \nr{nr} indicate that at $\rmO(M^0)$, 
$S = \theta^\dagger \theta - \phi^\dagger \phi$, which does not lead to
any non-trivial $t$-dependence. The leading non-trivial term reads
\ba
 S(x;\vec{r}) & = &  
 ... + \frac{i}{2M} 
 \biggl[ 
 \theta^\dagger\Bigl( t,\vec{x}+ \frac{\vec{r}}{2} \Bigr)
 \overleftrightarrow{\!D}_{\!\!j}\sigma_j 
 \phi \Bigl( t,\vec{x} - \frac{\vec{r}}{2} \Bigr)
 + \nn &  & \hspace*{1.35cm} +  
 \phi^\dagger \Bigl( t,\vec{x}+ \frac{\vec{r}}{2} \Bigr)
 \overleftrightarrow{\!D}_{\!\!j} \sigma_j 
 \theta \Bigl( t,\vec{x} - \frac{\vec{r}}{2} \Bigr)
 \biggr] + 
 \rmO\biggl( \frac{1}{M^2} \biggr)
 \;, \la{S_NRQCD}
\ea
where 
$
 %\stackrel{\leftrightarrow}{D_j} 
 \overleftrightarrow{\!D}_{\!\!j} 
 \; \equiv \; 
 W \! \overrightarrow{\!D}_{\!\!j}\! ( t,\vec{x} - {\vec{r}} / {2})\; - 
 \overleftarrow{\!D}_{\!\!j}\! ( t,\vec{x} + {\vec{r}} / {2})\, W
$.

%\newpage

To simplify \eq\nr{S_NRQCD} a bit, let us for now assume that the gauge fields
are perturbative, so that the Wilson line can be approximated by
the first term in its expansion; and that their variation is slow
on the scale set by $|\vec{r}|$, as argued in \se\ref{se:power}
(in any case, $|\vec{r}|$ is taken to be zero at the end).
Then we may write 
$
 W \approx \unit + i g \vec{r} \cdot \vec{A}(t,\vec{x})
$, 
$
 %\stackrel{\rightarrow}{D_j}\! 
 \overrightarrow{\!D}_{\!\!j}\! 
 ( t,\vec{x} - {\vec{r}} / {2})\! 
 \;\approx\; 
 \overrightarrow{\!\partial}_{\!\!j}\! - i g A_j(t,\vec{x}) 
 + ig \vec{r}\cdot \nabla A_j(t,\vec{x}) / 2
$, 
$
 \overleftarrow{\!D}_{\!\!j}\! ( t,\vec{x} + {\vec{r}} / {2})\!
 \;\approx\; 
 \overleftarrow{\!\partial}_{\!\!j}\! + i g A_j(t,\vec{x}) 
 + ig \vec{r}\cdot \nabla A_j(t,\vec{x}) / 2
$.
We now note that 
\be
 \theta^\dagger \Bigl( t,\vec{x}+ \frac{\vec{r}}{2} \Bigr) 
 \overleftrightarrow{\!D}_{\!\!j}
 \phi \Bigl( t,\vec{x} - \frac{\vec{r}}{2} \Bigr)
 \simeq - 2
 \frac{\partial}{\partial r^j} \Bigl\{ 
 \theta^\dagger \Bigl( t,\vec{x}+ \frac{\vec{r}}{2} \Bigr) W 
 \phi \Bigl( t,\vec{x} - \frac{\vec{r}}{2} \Bigr)
 \Bigr\} 
 \;. \la{CVCS_rel_pre}
\ee
Therefore, to leading order in the large-$M$ expansion, and at least
to some order in the weak-coupling expansion, we can identify
\be
 S(x;\vec{r}) \simeq -\frac{i}{M} \nabla_{\vec{r}} \cdot 
 \vec{V}(x;\vec{r})
 \;,  \la{SVrel}
\ee
where the components of $\vec{V}$ are given in \eq\nr{Vk_nr}. 
% In other words, if $\vec{V}(x;\vec{r})$ represents 
% an S-wave state, then $S(x;\vec{r})$ has the quantum numbers 
% of a {\em P-wave} state. 

The relation between the vector and scalar channel correlators can 
be pushed one step further, if we consider directly the correlators, 
\eqs\nr{CV} and \nr{CS}. To leading order in the large-$M$ expansion,
\eqs\nr{Vk_nr} and \nr{SVrel} imply that 
\ba
 C_{>}^{S}(x;\vec{r},\vec{r}') 
 & = &  
 \Bigl\langle
   S (x;\vec{r}) S (0;-\vec{r}')
 \Bigr\rangle
 \nn & \simeq & 
 \frac{1}{M^2} \sum_{k,l=1}^{3}
 (\nabla_\vec{r})_k(\nabla_\vec{r'})_l
 \Bigl\langle
   V_k (x;\vec{r}) V_l (0;-\vec{r}')
 \Bigr\rangle
 \nn & = & 
 \frac{1}{3 M^2} \sum_{k,l=1}^{3}
 (\nabla_\vec{r})_k(\nabla_\vec{r'})_l \; \delta_{kl}
 \sum_{j=1}^{3} \Bigl\langle
   V_j (x;\vec{r}) V_j (0;-\vec{r}')
 \Bigr\rangle
 \nn & = & 
 -\frac{1}{3 M^2} \nabla_\vec{r}\cdot \nabla_{\vec{r}'} 
 \sum_{j=1}^{3} \Bigl\langle
   V^j (x;\vec{r}) V_j (0;-\vec{r}')
 \Bigr\rangle
 \nn & = & 
 -\frac{1}{3 M^2} \nabla_\vec{r}\cdot \nabla_{\vec{r}'} \; 
 C_{>}^{V}(x;\vec{r},\vec{r}')
 \;. \la{CVCS_rel}
\ea
We will be making use of this important relation later on.  

% \subsubsection{Tree-level computation}

We now return to full QCD, 
and outline the computation of the 2-point
scalar density correlator in \eq\nr{CS} at tree-level, 
again taking a spatial Fourier transfrom and,  
for generality, keeping track
of a non-zero spatial momentum $\vec{q}\neq\vec{0}$ for the moment. Then, 
\ba
 & & \hspace*{-1cm} C_E^S(\tau,\vec{q};\vec{r,r'}) \nn
 & \equiv & 
 \int \! {\rm d}^3 \vec{x}\, e^{-i \vec{q}\cdot \vec{x}}
 \Bigl\langle
  {\!\bar\psi}\,\Bigl(\tau,\vec{x} +\frac{\vec{r}}{2}
  \Bigr)
  %%  \gamma^\mu
  \, % W_{\vec{r}}[(t,\vec{x}+\frac{\vec{r}}{2});
     %                    (t,\vec{x}-\frac{\vec{r}}{2})] \, 
  \psi\Bigl(\tau,\vec{x} -\frac{\vec{r}}{2}
  \Bigr) \;\; 
  {\!\bar\psi}\,\Bigl(0,- \frac{\vec{r'}}{2}\Bigr)
  %%  \gamma_\mu
  {\psi}\Bigl(0,+ \frac{\vec{r'}}{2}\Bigr)
 \Bigr\rangle
 \la{Ppoinsplit} \\[2mm]
  &  = & 
 - \Nc \int \! {\rm d}^3 \vec{x} \, e^{-i \vec{q}\cdot \vec{x}}
   \Tint{\tilde P_\fe, \tilde S_\fe}
   e^{i (\tilde p_\zfe - \tilde s_\zfe)\tau + i (\vec{s}-\vec{p})\cdot \vec{x}
                    + i (\vec{s}+\vec{p})\cdot \frac{\vec{r-r'}}{2}
     }
   \tr
   \biggl[
    %% \tilde \gamma_\mu
     \frac{-i \bsl{\tilde P} + M }{\tilde P^2 + M^2} 
    %% \tilde \gamma_\mu 
     \frac{-i \bsl{\tilde S} + M }{\tilde S^2 + M^2} 
   \biggr]   
 \nn[2mm] & = & 
 - 4 \Nc 
   \Tint{\tilde P_\fe, \tilde S_\fe}
   (2\pi)^3 \delta^{(3)}(\vec{s} - \vec{p}-\vec{q})
   e^{i (\tilde p_\zfe - \tilde s_\zfe)\tau 
      + i (2 \vec{p} + \vec{q}) \cdot \frac{\vec{r-r'}}{2}
   }
   \frac{- \tilde p_\zfe \tilde s_\zfe - \vec{p}\cdot \vec{s} + M^2 }
        {(\tilde P^2 + M^2)(\tilde S^2 + M^2)}
  \nn[2mm] & = & 
 - 2 \Nc 
   T^2 \sum_{\tilde p_\zfe, \tilde s_\zfe} \int \! 
   \frac{{\rm d}^3 \vec{p}}{(2\pi)^3} 
   e^{i (\tilde p_\zfe - \tilde s_\zfe)\tau 
      + i (2 \vec{p} + \vec{q}) \cdot \frac{\vec{r-r'}}{2}
   }
   \frac{
  - 2 \tilde p_\zfe \tilde s_\zfe 
  - E_\vec{p}^2 - E_\vec{p+q}^2 + 4 M^2 + \vec{q}^2 
        }
        {(\tilde p_\zfe^2 + E_\vec{p}^2)
         (\tilde s_\zfe^2 + E_\vec{p+q}^2)}
%  \times \nn & & \times  
%  \Bigl( 
%  \Bigr)
   \;. \nn
\ea
Making use of \eqs\nr{Msum1}, \nr{Msum2}, this can be rewritten as
\ba
 && \!\!\!\! C_E^S(\tau,\vec{q};\vec{r,r'}) = 
 - \Nc \int \! \frac{{\rm d}^3\vec{p}}{(2\pi)^3} 
  e^{i (2 \vec{p} + \vec{q}) \cdot \frac{\vec{r-r'}}{2}}
 \times \nn && \times \biggl\{  
%%%
 \nF{}(E_\vec{p}+\mu) \nF{}(E_\vec{p+q}-\mu)
 e^{(\beta-\tau)(E_\vec{p}+E_\vec{p+q})}
 \biggl[ 
% -1 + \frac{\vec{q}^2 + 4 M^2 - E_\vec{p}^2 - E_\vec{p+q}^2}
%          {2 E_\vec{p}E_\vec{p+q}} 
 \frac{\vec{q}^2 + 4 M^2 - (E_\vec{p} + E_\vec{p+q})^2}
          {2 E_\vec{p}E_\vec{p+q}} 
 \biggr] 
 + \nn && +     
%%%
  \nF{}(E_\vec{p}+\mu) \nF{}(E_\vec{p+q}+\mu)
 e^{(\beta-\tau)E_\vec{p}+\tau E_\vec{p+q}+\beta\mu}
 \biggl[ 
% -1 - \frac{\vec{q}^2 + 4 M^2 - E_\vec{p}^2 - E_\vec{p+q}^2}
%          {2 E_\vec{p}E_\vec{p+q}} 
 \frac{-\vec{q}^2 - 4 M^2 +( E_\vec{p} - E_\vec{p+q})^2}
          {2 E_\vec{p}E_\vec{p+q}} 
 \biggr] 
 + \nn && +     
%%%
  \nF{}(E_\vec{p}-\mu) \nF{}(E_\vec{p+q}-\mu)
 e^{(\beta-\tau)E_\vec{p+q}+\tau E_\vec{p}-\beta\mu}
 \biggl[ 
% -1 - \frac{\vec{q}^2 + 4 M^2 - E_\vec{p}^2 - E_\vec{p+q}^2}
%          {2 E_\vec{p}E_\vec{p+q}} 
 \frac{-\vec{q}^2 - 4 M^2 +( E_\vec{p} - E_\vec{p+q})^2}
          {2 E_\vec{p}E_\vec{p+q}} 
 \biggr] 
 + \nn && +     
%%%
 \nF{}(E_\vec{p}-\mu) \nF{}(E_\vec{p+q}+\mu)
 e^{\tau(E_\vec{p}+E_\vec{p+q})}
 \biggl[ 
% -1 + \frac{\vec{q}^2 + 4 M^2 - E_\vec{p}^2 - E_\vec{p+q}^2}
%          {2 E_\vec{p}E_\vec{p+q}} 
 \frac{\vec{q}^2 + 4 M^2 - (E_\vec{p} + E_\vec{p+q})^2}
          {2 E_\vec{p}E_\vec{p+q}} 
 \biggr] 
 \biggr\}
 \;. \la{PqcdCE}
\ea
With the same considerations as between \eqs\nr{SqcdCE} and \nr{Ssimple}, 
the interesting part of $C_E^S$ can be approximated as
\ba
 && \hspace*{-2cm} C_E^S(\tau,\vec{q};\vec{r,r'}) \simeq
 \frac{\Nc}{2 M^2} \int \! \frac{{\rm d}^3\vec{p}}{(2\pi)^3} 
  e^{i (2 \vec{p} + \vec{q}) \cdot \frac{\vec{r-r'}}{2} %%}
% \times \nn && \times \biggl\{  
%%%
 %% \nF{}(E_\vec{p}+\mu) \nF{}(E_\vec{p+q}-\mu)
%% e^{
  -\tau \bigl[ 2 M + 
 \frac{2 \vec{p}^2 + 2 \vec{p}\cdot \vec{q} + \vec{q}^2}{2 M}
 + \rmO\bigl( \frac{1}{M^3} \bigr) \bigr]}
 \times \nn && \times 
 \biggl[ 
  4 \vec{p}^2 + 4 \vec{p}\cdot \vec{q} + \vec{q}^2
 + \rmO\biggl( \frac{1}{M^2} \biggr)
 \biggr] 
 %% \biggr\}
 \;. \la{Psimple}
\ea
Note again that after these simplifications, all dependence on the 
temperature and on the chemical potential 
has disappeared from the tree-level result. 

% \subsubsection{Representation through a Schr\"odinger-equation}

The exponential in \eq\nr{Psimple} is the same as in \eq\nr{Ssimple}, 
whereby $C_{>}^S$ obeys the same Schr\"odinger equation as $C_{>}^V$, 
\eq\nr{SSchr}. The initial condition is different, however:
setting $t=0$ in \eq\nr{Psimple} (after $\tau\to it$), 
we find
\ba
 C_{>}^S(0,\vec{q};\vec{r,r'}) & = &  - \frac{ 2 \Nc }{M^2}
 \nabla_\vec{r}^2 \int \! \frac{{\rm d}^3\vec{p}}{(2\pi)^3} 
 e^{i (2 \vec{p} + \vec{q}) \cdot \frac{\vec{r-r'}}{2} }
 + \rmO\biggl(\frac{1}{M^4} \biggr)
 \nn & = & 
 - \frac{ 2 \Nc }{M^2}
 \nabla_\vec{r}^2 \,  \delta^{(3)}(\vec{r-r'})
 + \rmO\biggl(\frac{1}{M^4} \biggr)
 \;.
 \la{Pinit0}
\ea
This agrees, of course, with what can be deduced from 
\eqs\nr{Sinit0}, \nr{CVCS_rel}.
We note that all dependence on the external momentum $\vec{q}$ again 
only appears as a part of the center-of-mass energy 
$
 2 M + \vec{q}^2/4 M
$, inside \eq\nr{SSchr}. 

For future reference, let us finally 
determine the spectral function, $\rho^S(\omega)$. 
\eq\nr{Psimple} (after $\tau\to it$) already
shows the solution of \eqs\nr{SSchr}, \nr{Pinit0}, 
and we can then directly remove the point-splitting, 
setting $\vec{r,r'} = \vec{0}$.  Shifting $\vec{p}\to \vec{p} - \vec{q}/2$; 
taking the steps in footnote~\ref{rhorec}; and ignoring exponentially
small terms, we find
\ba
 \rho^S(\omega) & \approx & \frac{2\Nc\pi}{M^2}
 \int \! \frac{{\rm d}^3\vec{p}}{(2\pi)^3}
 \, \vec{p}^2 \, 
 \delta\Bigl( \omega' - \frac{\vec{p}^2}{M}\Bigr)
 = % \nn & = & 
 \frac{\Nc}{2\pi} \, \theta(\omega')\, M^{\fr12} (\omega')^{\fr32}
 \;, \la{tree_P}
\ea
where $\omega'$ is from \eq\nr{omegap}.

The analysis so far has been at tree-level. As discussed 
above \eq\nr{CVCS_rel_pre}, the relation in~\eq\nr{CVCS_rel} 
is more general, however. Therefore, we can extract a beyond-the-leading
order $\rho^S$ by simply applying \eq\nr{CVCS_rel} to a beyond-the-leading
order $\rho^V$.

%%%%%%%%%%%%%%%%%%%%%% SUBSECTION %%%%%%%%%%%%%%%%%%%%%%%%%%%%%%%%%%%%%
%
\subsection{Other channels}

In Secs.~\ref{ss:rhoV}, \ref{ss:rhoS}, 
we have discussed the correlators in the 
vector and scalar channels. Let us now show that in the limit of a large
quark mass, the correlators in the pseudoscalar and axial vector channels 
are to a good approximation equivalent to either of these two. 

We note, first of all, that in the basis of \eq\nr{Dirac_NRQCD}, the
matrix $\gamma_5$ becomes 
\be
 \gamma_5 = i \gamma^0\gamma^1\gamma^2\gamma^3 = 
  \left( 
  \begin{array}{cc}
   0 & \unit  \\   
   \unit & 0  
  \end{array}
 \right)
 \;. \la{gamma5}
\ee
Thereby the pseudoscalar density becomes
\ba
 P(x;\vec{r}) & \equiv &  
 \bar\psi\Bigl(t,\vec{x}+\frac{\vec{r}}{2}\Bigr) 
 i \gamma_5  \, W
 \psi\Bigl(t,\vec{x}-\frac{\vec{r}}{2}\Bigr)
 \la{Pdef} \\ 
 & = & 
 i \Bigl[ \theta^\dagger\Bigl( t,\vec{x}+ \frac{\vec{r}}{2} \Bigr)
 \, W 
 \phi \Bigl( t,\vec{x} - \frac{\vec{r}}{2} \Bigr)
 - 
 \phi^\dagger\Bigl( t,\vec{x}+ \frac{\vec{r}}{2} \Bigr)
 \, W 
 \theta \Bigl( t,\vec{x} - \frac{\vec{r}}{2} \Bigr) \Bigr] 
 + \rmO\Bigl( \frac{1}{M^2} \Bigr)
 \;, \nn \la{P_nr}
\ea
where again only structures of the type 
$\theta^\dagger\phi$ and $\phi^\dagger\theta$ have been kept. 
The non-trivial two-point correlator comes from the cross-term
between the two structures in \eq\nr{P_nr}, and ignoring the 
spin-dependent corrections of $\rmO(1/M)$, a comparison with 
\eq\nr{Vk_nr} then shows that 
\be
 C_{>}^{P}(x;\vec{r},\vec{r}') \simeq -\fr13 C_{>}^{V}(x;\vec{r},\vec{r}')
 \;,  \la{CP}
\ee
where 
$
 C_{>}^{P}(x;\vec{r},\vec{r}') \equiv
 \langle 
   P(x;\vec{r}) P(0;-\vec{r}')
 \rangle 
$, 
and $C_{>}^{V}$ is defined in \eq\nr{CV}.

The axial vector, on the other hand, can be defined as
\be
 A^\mu(x;\vec{r}) \equiv 
 \bar\psi\Bigl(t,\vec{x}+\frac{\vec{r}}{2}\Bigr) 
 \gamma_5 \gamma^\mu  \, W
 \psi\Bigl(t,\vec{x}-\frac{\vec{r}}{2}\Bigr)
 \;.  \la{Amudef}
\ee
In the case of $V^\mu$, we found that the dominant contribution is 
given by the spatial components, but for the axial vector, the roles 
have interchanged: the leading term is
\be
 A^0(x;\vec{r}) = 
 - \Bigl[ \theta^\dagger\Bigl( t,\vec{x}+ \frac{\vec{r}}{2} \Bigr)
 \, W 
 \phi \Bigl( t,\vec{x} - \frac{\vec{r}}{2} \Bigr)
 + 
 \phi^\dagger\Bigl( t,\vec{x}+ \frac{\vec{r}}{2} \Bigr)
 \, W 
 \theta \Bigl( t,\vec{x} - \frac{\vec{r}}{2} \Bigr) \Bigr] 
 + \rmO\Bigl( \frac{1}{M^2} \Bigr)
 \;. \nn \la{A0_nr}
\ee
Comparing with \eq\nr{P_nr}, we find 
\be
 C_{>}^{A^0}(x;\vec{r},\vec{r}') \equiv
 \langle 
   A^0(x;\vec{r}) A^0(0;-\vec{r}')
 \rangle 
 \simeq  C_{>}^{P}(x;\vec{r},\vec{r}')
 \simeq -\fr13 C_{>}^{V}(x;\vec{r},\vec{r}') 
 \;. \la{CA0}
\ee
In lattice studies, however, attention is sometimes restricted
to the spatial components $A^k$; repeating the previous steps, we find 
\ba
 A^k(x;\vec{r})  & \simeq & 
 -\frac{1}{2M} \frac{\partial}{\partial x^k} P(x,\vec{r})   
 + \nn   &  & \hspace*{-1.5cm} + 
 \frac{\epsilon_{klm}}{M}  \frac{\partial}{\partial r^m}
 \Bigl[ \theta^\dagger\Bigl( t,\vec{x}+ \frac{\vec{r}}{2} \Bigr)
 \sigma_l \, W 
 \phi \Bigl( t,\vec{x} - \frac{\vec{r}}{2} \Bigr)
 - 
 \phi^\dagger\Bigl( t,\vec{x}+ \frac{\vec{r}}{2} \Bigr)
 \sigma_l \, W 
 \theta \Bigl( t,\vec{x} - \frac{\vec{r}}{2} \Bigr)
 \Bigr]
 \;. \hspace*{0.5cm} \la{Ak_nr}
\ea
The first term is a total derivative, and the second term has
a structure close to that in \eq\nr{Vk_nr}, given that
only the crossterm contributes in a correlation function. 
Therefore, paralleling the argument in \eq\nr{CVCS_rel}, we find
\ba
% & & \hspace*{-1cm}
% \left. 
 \int\! {\rm d}^3\vec{x} \,
 C_{>}^{\vec{A}}(x;\vec{r,r'}) & \equiv & 
 \int\! {\rm d}^3\vec{x} \,
 \langle 
   A^k(x;\vec{r}) A^k (0;-\vec{r}')
 \rangle
% \right|_\rmi{no sum over $k$}	 
 \nn 
  & \simeq &  
 \frac{1}{M^2}
% \left.
 \epsilon_{klm}\epsilon_{kl'm'}
 \frac{\partial^2}{\partial r^m \partial r'^{m'}}
 \int\! {\rm d}^3\vec{x} \,
 \langle 
     V^l (x;\vec{r}) V^{l'} (0;-\vec{r}')
 \rangle
% \right|_\rmi{no sum over $k$}
 \nn  & = &
 -\frac{1}{3 M^2}
% \left.
 \epsilon_{klm}\epsilon_{kl'm'}
 \frac{\partial^2}{\partial r^m \partial r'^{m'}} \delta_{ll'}
 \int\! {\rm d}^3\vec{x} \,
 C_{>}^{V}(x;\vec{r},\vec{r}')
% \right|_\rmi{no sum over $k$}
 \nn  & = &  	 
 2  \int\! {\rm d}^3\vec{x} \,
 C_{>}^{S}(x;\vec{r},\vec{r}')
 \;. \la{CACS_rel}
\ea
To summarize, \eqs\nr{CP}, \nr{CA0} and \nr{CACS_rel}
show that the pseudoscalar and axial correlators do not
lead to any qualitatively new structures. 

%%%%%%%%%%%%%%%%%%%%%%%%% SECTION %%%%%%%%%%%%%%%%%%%%%%%%%%%%%%%%%%%%%
%
\section{Method to construct the spectral functions}
\la{se:rho}

In the previous section, we have set up the Schr\"odinger 
equation and initial conditions satisfied by the vector
channel correlator $C_{>}^{V}$, and shown that the corresponding
correlators in the other channels can be obtained from $C_{>}^{V}$
through various relations. The aim now is to extract the spectral
functions corresponding to these correlators.  

To achieve this goal, it is useful to convert the time-dependent
Schr\"odinger equation directly to frequency space. Let 
\be
 \psi(t;\vec{r,r'}) \equiv 
 e^{i 2 M t}\, C_{>}^{V}(t,\vec{q};\vec{r,r'})
 \;, \la{psi_def}
\ee 
and
\be
 \chi(t;\vec{r,r'}) \equiv e^{i 2 M t} C_{>}^{S}(t,\vec{q};\vec{r,r'})
 \simeq - \fr13 
 \frac{\nabla_\vec{r}\cdot \nabla_{\vec{r}'}}{M^2}
 \psi(t;\vec{r,r'})
 \;. \la{chi}
\ee
The corresponding frequency representations
are defined by
\be
 \tilde\psi(\omega';\vec{r,r'}) \equiv 
 \int_{-\infty}^{\infty}
 \! {\rm d}t \, e^{i \omega' t} \, \psi(t;\vec{r,r'})
 \;, \quad
 \tilde\chi(\omega';\vec{r,r'}) \equiv
 \int_{-\infty}^{\infty}
 \! {\rm d}t \, e^{i \omega' t} \, \chi(t;\vec{r,r'})
 \;,
\ee
and the spectral functions are then obtained from 
(cf.\ \eq\nr{rhoV})
\ba
 \rho^V(\omega') & = &  \lim_{\vec{r,r'}\to \vec{0}} 
 \fr12 \tilde \psi(\omega';\vec{r,r'}) 
 \;, \la{rhoV_Psi_pre}\\
 \rho^S(\omega') & = & 
 \lim_{\vec{r,r'}\to \vec{0}} 
 \fr12
 \tilde \chi(\omega';\vec{r,r'})
 \;, \la{rhoS_Psi_pre}
\ea
where $\omega'$ is from \eq\nr{omegap} and
we have omitted exponentially small corrections. 

We now recall from ref.~\cite{static} that 
the imaginary part of $V_>(t,r)$ (\eq\nr{expl})
is odd in $t\to -t$. Furthermore, we recall from \se\ref{se:power} that
a consistent perturbative solution allows (or, to be more precise, demands)
considering the limit $|t| \gg r$. Denoting 
\be
 V_{>}(r) \equiv \lim_{t\to + \infty} V_{>}(t,r)
 \;,  
\ee
the equations to be solved (\eq\nr{Seq}) then read
\ba
 \Bigl[ \hat H - i |\im V_{>}(r)| \Bigr]
 \psi(t;\vec{r,r'}) & = &  
 i \partial_t \psi(t;\vec{r,r'}) \;, \quad t > 0 
 \;, \la{Spsia} \\  
 \Bigl[ \hat H + i |\im V_{>}(r)| \Bigr]
 \psi(t;\vec{r,r'}) & = &  
 i \partial_t \psi(t;\vec{r,r'}) \;, \quad t < 0 
 \;, \la{Spsib}
\ea
where we indicated explicitly that the imaginary part is 
negative for $t\to+ \infty$~\cite{static,imV}, and defined
a Hermitean differential operator $\hat H$ through 
\be
 \hat H \equiv - \frac{\nabla_\vec{r}^2}{M}
 + \re V_{>}(r)
 \;.
\ee

Since the effective Hamiltonian is time-independent 
both for $t<0$ and for $t>0$, we can  formally solve 
\eqs\nr{Spsia}, \nr{Spsib}: 
\be
 \psi(t;\vec{r,r'}) = 
 \left\{ 
 \begin{array}[c]{ll}
   e^{-i \hat H t - |\rmi{Im}\, V_{>}(r)| t}\, 
   \psi(0;\vec{r,r'}) & \;, \quad t > 0 \\ 
   e^{-i \hat H t + |\rmi{Im}\, V_{>}(r)| t}\, 
   \psi(0;\vec{r,r'}) & \;, \quad t < 0 \\ 
 \end{array}
 \right.
 \;,
\ee
where, according to \eqs\nr{Sinit0}, \nr{psi_def},  
\be
 \psi(0;\vec{r,r'}) = - 6 \Nc \delta^{(3)}(\vec{r-r'})
 \;.
 \la{psi_init}
\ee
Taking a Fourier-transform, we get
\ba
 \tilde \psi(\omega';\vec{r,r'}) & = & 
 \int_{-\infty}^{\infty}
 \! {\rm d}t \, 
 e^{i\omega' t} \psi(t;\vec{r,r'})
 \nn & = &  
 \biggl\{ 
 \Bigl[ i\omega' - i \hat H - |\im V_{>}(r)|\Bigr]^{-1}
 \left. e^{it(\omega' - \hat H) - |\rmi{Im}\, V_{>}(r)| t} 
 \right|^{\infty}_{0}
 + \nn  & &  \hspace*{0.0cm} + 
 \Bigl[ i\omega' - i \hat H + |\im V_{>}(r)|\Bigr]^{-1}
 \left. e^{it(\omega' - \hat H) + |\rmi{Im}\, V_{>}(r)| t} 
 \right|^{0}_{-\infty}
 \biggr\}\;\psi(0;\vec{r,r'})
 \nn & = & 
\frac{1}{i}
 \biggl\{ 
 \Bigl[ \omega' - \hat H - i |\im V_{>}(r)|\Bigr]^{-1}
  -  % \nn  & &  \hspace*{0.3cm} -
 \Bigl[ \omega' - \hat H + i |\im V_{>}(r)|\Bigr]^{-1}
 \biggr\}\; \psi(0;\vec{r,r'})
 \;. \nn  \la{Seq_omega_pre}
\ea
To give a concrete meaning to the inverses
in \eq\nr{Seq_omega_pre}, we define 
a function $\tilde \Psi(\omega';\vec{r,r'})$ as the solution 
of the equation
\be
 \Bigl[
   \omega' - \hat H + i |\im V_{>}(r)|  
 \Bigr]
 \tilde \Psi(\omega';\vec{r,r'}) = - 6 \Nc \delta^{(3)}(\vec{r-r'})
 \;. \la{Seq_omega}
\ee
Then the result of \eq\nr{Seq_omega_pre} can be rewritten as
\be
 \tilde \psi(\omega';\vec{r,r'}) = 
 - 2 \im \Bigl[ \tilde \Psi(\omega';\vec{r,r'}) \Bigr] 
 \;. \la{tildepsi_res}
\ee
According to \eqs\nr{chi}, \nr{rhoV_Psi_pre}, \nr{rhoS_Psi_pre}, 
the spectral functions are now obtained from 
\ba
 \rho^V(\omega') & = &  
% \lim_{\vec{r,r'}\to \vec{0}} 
% \fr12 \tilde \psi(\omega';\vec{r,r'}) 
% = 
 - 
 \lim_{\vec{r,r'}\to \vec{0}} 
 \im \Bigl[ \tilde \Psi(\omega';\vec{r,r'}) \Bigr]
 \;, \la{rhoV_Psi}\\
 \rho^S(\omega') & \simeq & 
% \lim_{\vec{r,r'}\to \vec{0}} 
% \fr12
% \tilde \chi(\omega';\vec{r,r'})
% = 
 \lim_{\vec{r,r'}\to \vec{0}} 
 \frac{1}{3 M^2}
 \im \Bigl[ \nabla_\vec{r}\cdot \nabla_{\vec{r}'} 
 \tilde \Psi(\omega';\vec{r,r'}) \Bigr]
 \;. \la{rhoS_Psi}
\ea
To summarize, we have reduced the determination of the spectral 
functions to the solution of a time-independent
inhomogeneous Schr\"odinger equation, \eq\nr{Seq_omega}. 

As the next step, following ref.~\cite{ps}, we introduce the ansatz
\be
 \tilde \Psi(\omega';\vec{r,r'})
 \equiv 
 \sum_{l=0}^{\infty} \sum_{m=-l}^{l}
 \frac{\tilde g_l(\omega';r,r')}{rr'}
 Y_{lm}(\Omega)Y_{lm}^*(\Omega')
 \;. \la{omega_rep}
\ee
Here $Y_{lm}$ are the spherical harmonics, normalised as
$
 \int \! {\rm d}\Omega \, Y^*_{lm}(\Omega) Y_{l'm'}(\Omega) 
 = \delta_{ll'} \delta_{mm'}
$,
where 
$
 {\rm d}\Omega = {\rm d}\!\cos\theta\, {\rm d}\phi
$, 
and satisfying
\be
 \sum_{lm} Y^*_{lm}(\Omega') Y_{lm}(\Omega) =
 \delta(\Omega - \Omega') \equiv 
 \delta(\cos\theta - \cos\theta') \, \delta(\phi-\phi') 
 \;.
 \la{spher_complete}
\ee
The $\delta$-function can be written as 
\be
 \delta^{(3)}(\vec{r} - \vec{r}') = 
 \frac{1}{rr'} \delta(r-r') \delta(\Omega - \Omega')
 \;, \la{spher_delta}
\ee
whereby \eq\nr{Seq_omega} becomes
\be
 \Bigl[
   \omega' - \hat H_r + i |\im V_{>}(r)|  
 \Bigr]
 \tilde g_l(\omega';r,r') = - 6 \Nc \delta(r-r')
 \;, \la{Seq_omega_l}
\ee
with 
\be
 \hat H_r \equiv -\frac{1}{M} \frac{\partial^2}{\partial r^2} 
 + \frac{l(l+1)}{Mr^2}
 + \re V_{>}(r)
 \;. \la{Hr}
\ee

The remaining goal is to reduce the problem to the solution of 
the homogeneous equation. 
Following refs.~\cite{ps,mp3}, we introduce the ansatz
\be
 \tilde g_l \equiv A \, g_{<}^l(r_{<}) g_{>}^l(r_{>})
 \;, \la{ansatz}
\ee
where
$g_{<}^l$ is a solution of the homogeneous equation regular at zero; 
$g_{>}^l$ is a solution of the homogeneous equation regular at infinity; 
and 
$
 r_{<} = \mathop{\mbox{min}}(r,r'), 
 r_{>} = \mathop{\mbox{max}}(r,r')
$.

Obviously, 
the function $\tilde g_l$ is symmetric in 
$r\leftrightarrow r'$, and continuous at $r=r'$. Given 
the well-known form of the solution $g_{<}^l$, it must thus behave as 
\be
 \tilde g_l \sim [ r^{l+1} + \rmO(r^{l+2})] [ (r')^{l+1} + \rmO((r')^{l+2}) ]
 \la{small_rrp}
\ee
at small $r,r'$. 
For the vector channel spectral function, \eqs\nr{rhoV_Psi}, \nr{omega_rep}
now imply that 
\be
  \rho^V(\omega') = 
 -  \lim_{r,r'\to 0} \frac{1}{4 \pi rr'}
 \im \Bigl[ \tilde g_0(\omega';r,r') \Bigr] 
 \;, \la{rho_res}
\ee
i.e.\ that only the S-wave ($l=0$) solution of the homogeneous 
part of \eq\nr{Seq_omega_l} contributes. 

Consider then the scalar channel. 
According to \eq\nr{rhoS_Psi}, the scalar channel spectral function
can be extracted from the same function $\tilde\Psi$ as the vector
channel one, by taking two derivatives 
and then extrapolating $r,r'\to 0$. Inspecting \eq\nr{small_rrp}, we
see that we at least get a contribution from the P-wave ($l=1$). 
However, according 
to \eq\nr{small_rrp}, it is also possible to get a contribution from the
{\em subleading S-wave terms}, 
$\tilde g_0 \sim [r+\rmO(r^2)][r'+\rmO((r')^2)]$. 
As far as we can see, this contribution
was omitted in ref.~\cite{mp3}.

We relegate a more detailed discussion on how to write
the solutions for the spectral functions $\rho^V, \rho^S$ to Appendix A, 
given that the further steps are quite technical in nature, 
and give here just the final formulae. Introducing the dimensionless
variables $\varrho\equiv r \alpha M$ and 
$\alpha \equiv g^2 C_F/4\pi$, the vector channel spectral 
function from \eq\nr{rho_res} can be simplified to 
\be
 \frac{\rho^V(\omega')}{M^2} = 
 - \frac{6 \Nc \alpha }{4\pi}
 \lim_{\delta\to 0} \int_{\delta}^{\infty} \! {\rm d}\varrho
 \im \left. \biggl\{ 
  \frac{1}{[g^0_{<}(\varrho)]^2} 
  \biggr\} \right|_{g^0_{<}(\varrho) = \varrho - \varrho^2/2 + ...}
 \;, \la{rho_res3}
\ee
while the scalar channel spectral function becomes
\be
 \frac{\rho^S(\omega')}{M^2}
 = 
 \frac{\Nc \alpha^3}{8\pi}  
 \lim_{\delta\to 0}
 \int_{\delta}^{\infty}
  \! {\rm d}\varrho \,
  \im \left. \biggl\{ 
  \frac{1}{[g^0_{<}(\varrho)]^2} 
 + 
  \frac{36}{[g^1_{<}(\varrho)]^2} 
  \biggr\} \right|_{g^1_{<}(\varrho) = \varrho^2 - \varrho^3/4 + ...}
 \;. \la{rhoS_full}
\ee
We remark that because of the factor 36, the first (S-wave) term is numerically
subdominant in \eq\nr{rhoS_full}, 
and would be totally negligible, were it not for the fact 
that is does lead to a resonance peak, unlike the second term.

%%%%%%%%%%%%%%%%%%%%%%%%% SECTION %%%%%%%%%%%%%%%%%%%%%%%%%%%%%%%%%%%%%
%
\section{Numerical results}
\la{se:num}

In the previous section, we have reduced 
the numerical determination of the vector and scalar channel
spectral functions to \eqs\nr{rho_res3}, \nr{rhoS_full}, 
respectively. In these equations the functions 
$g_{<}^{l}$, $l=0,1$, denote the regular solutions
of the homogeneous part of \eq\nr{Seq_omega_l}, 
\be
 \Bigl[
   \omega' - \hat H_r + i |\im V_{>}(r)|  
 \Bigr]
 g_{<}^l = 0
 \;, \la{Seq_omega_l_hom}
\ee
where $\hat H_r$ is from \eq\nr{Hr}.
Further details can be found in Appendix~A.

In practice, the procedure of determining
$\rho^V$, $\rho^S$ starts from some
small value, $\varrho\equiv \delta$, 
with for instance $\delta = 10^{-2}$, 
at which point we impose as initial conditions
the properties of the regular solutions at small $\varrho$, 
 $g^0_{<}(\delta) = \delta - \delta^2/2 + ...$~, 
 $g^1_{<}(\delta) = \delta^2 - \delta^3/4 + ...$~. 
We then integrate 
\eq\nr{Seq_omega_l_hom} towards larger $\varrho$, 
constructing simultaneously the quantities in 
\eqs\nr{rho_res3}, \nr{rhoS_full}. After a while,  
$g^0_{<}(\varrho)$ and $g^1_{<}(\varrho)$ start to 
grow rapidly and the integrals in \eqs\nr{rho_res3}, \nr{rhoS_full}
settle to their asymptotic values. Subsequently, we check that the
results obtained are independent of the starting point $\delta$. 
The numerics is straightforward and poses no problems. 

Apart from the pole mass $M$, 
the solution depends on what is plugged in for $g^2$ and 
$\mD$. We employ here simple analytic expressions 
that can be extracted from Ref.~\cite{adjoint}, 
\be
 g^2 \simeq \frac{8\pi^2}{9 \ln( 9.082\, T/ \Lambdamsbar)} 
 \;, \quad
 \mD^2 \simeq \frac{4\pi^2 T^2}{3 \ln(7.547\, T/\Lambdamsbar)}
 \;,
 \qquad \mbox{for $\Nc = \Nf = 3, \mu = 0$}
 \;. \la{numg2}
\ee 
We also fix $\Lambdamsbar \simeq 300$~MeV; for the uncertainties 
related to this, see \fig2 of ref.~\cite{static}.

%%%%%%%%%%%%%%%%%%%%%%%%% SUBSECTION %%%%%%%%%%%%%%%%%%%%%%%%%%%%%%%%%%
%
%\subsection{S-wave}

%%%%%%%%%%%%%%%%%%%%%%%%%%%%%%%%% FIGURE %%%%%%%%%%%%%%%%%%%%%%%%%%%%%%%%%
\begin{figure}[tb]

\centerline{%
\epsfysize=5.0cm\epsfbox{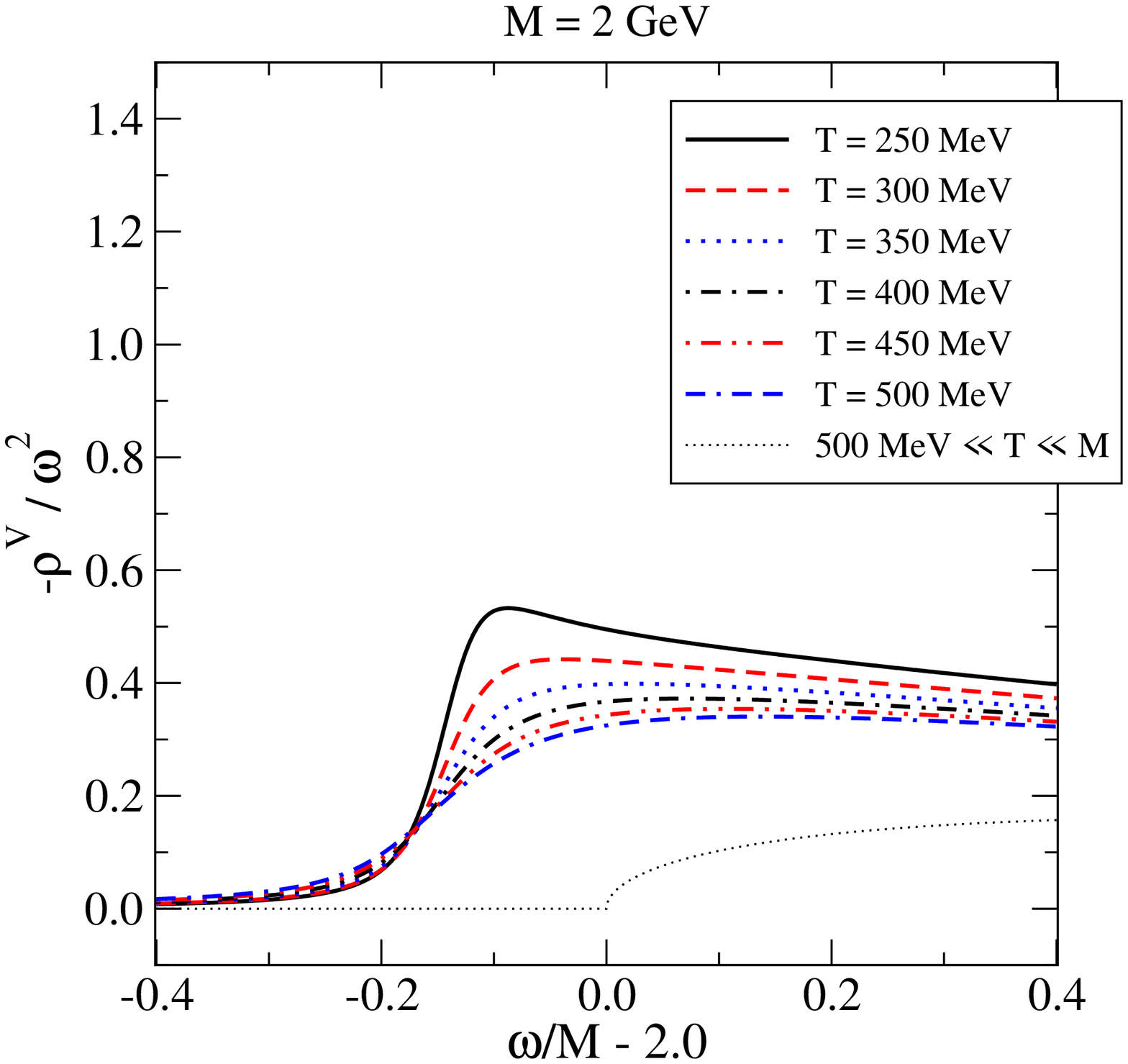}%
~~\epsfysize=5.0cm\epsfbox{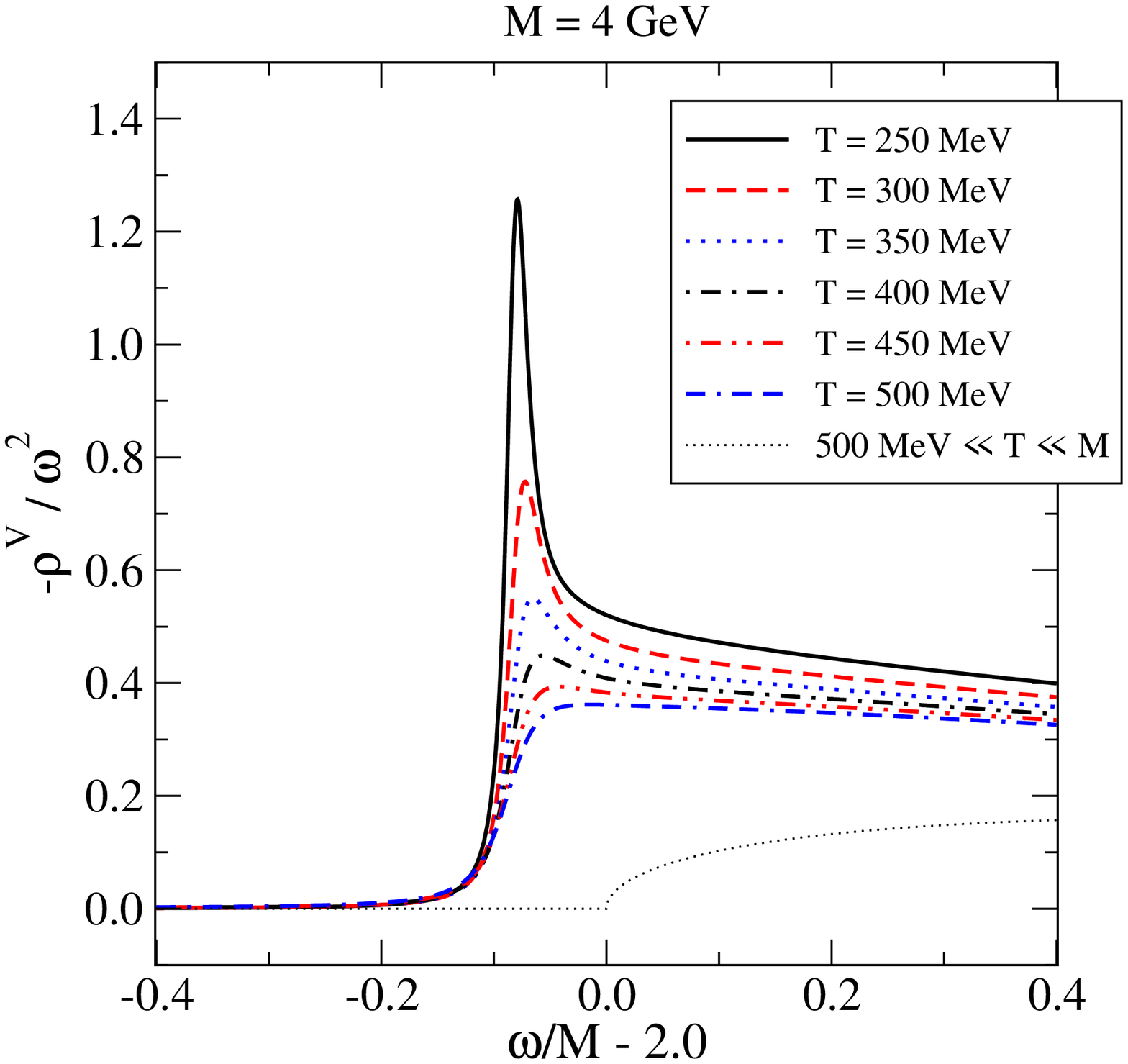}%
~~\epsfysize=5.0cm\epsfbox{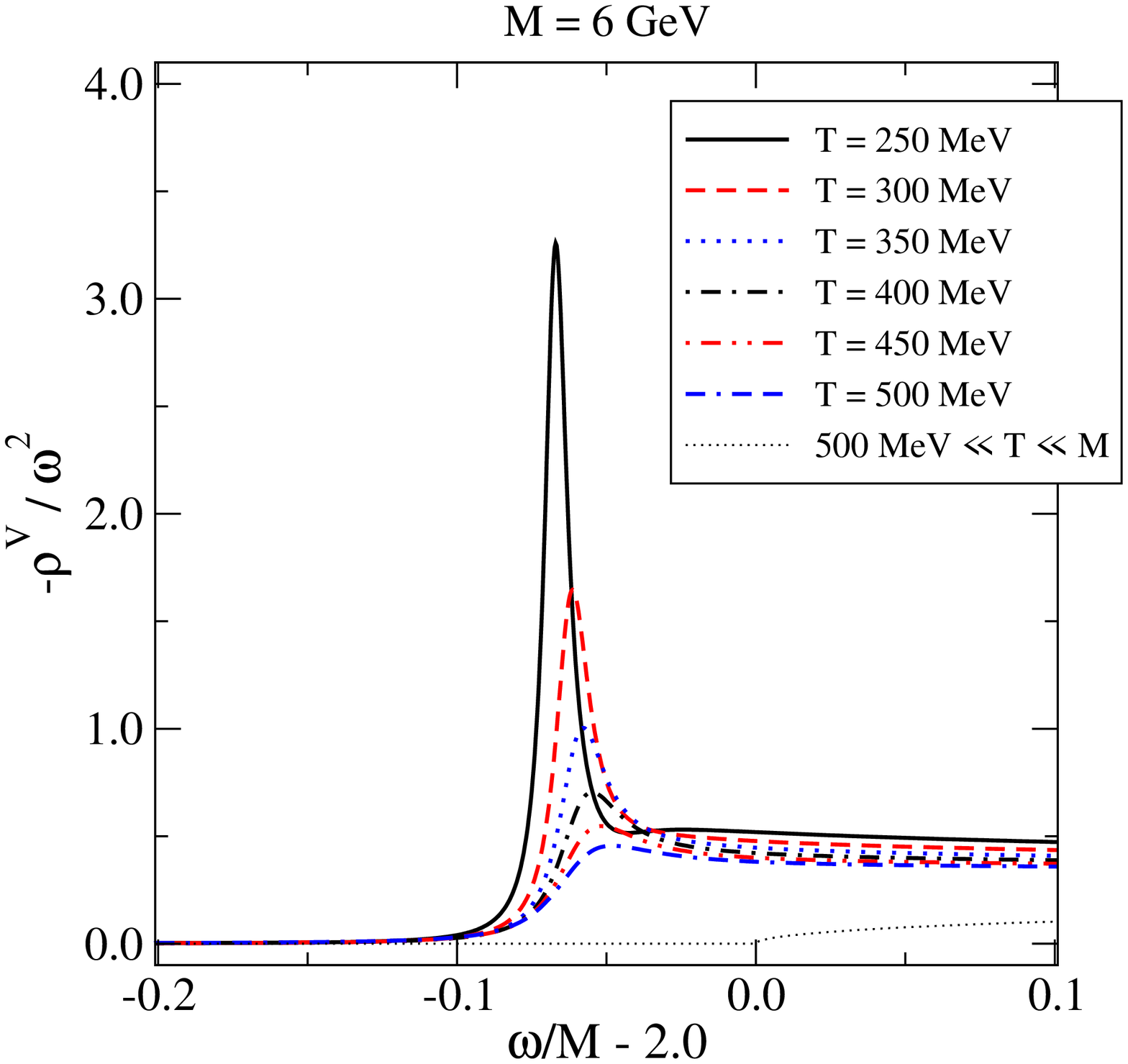}%
}

\vspace*{0.5cm}

%\vspace*{-4cm}

\caption[a]{The resummed perturbative vector channel
spectral function $\rho^V(\omega)$, in units of $\omega^2$, 
in the non-relativistic regime,
$(\omega - 2 M) / M \ll 1$, for $M = 2, 4, 6$~GeV (from left to right).
To the order considered, $M$ is the heavy quark pole mass. Note that
for better visibility, the axis ranges are different in the rightmost 
figure. 
}

\la{fig:rhoV_M}
\end{figure}
%%%%%%%%%%%%%%%%%%%%%%%%%%%%%%%%%%%%%%%%%%%%%%%%%%%%%%%%%%%%%%%%%%%%%%%%%%%

%%%%%%%%%%%%%%%%%%%%%%%%%%%%%%%%% FIGURE %%%%%%%%%%%%%%%%%%%%%%%%%%%%%%%%%
\begin{figure}[tb]

\centerline{%
\epsfysize=5.0cm\epsfbox{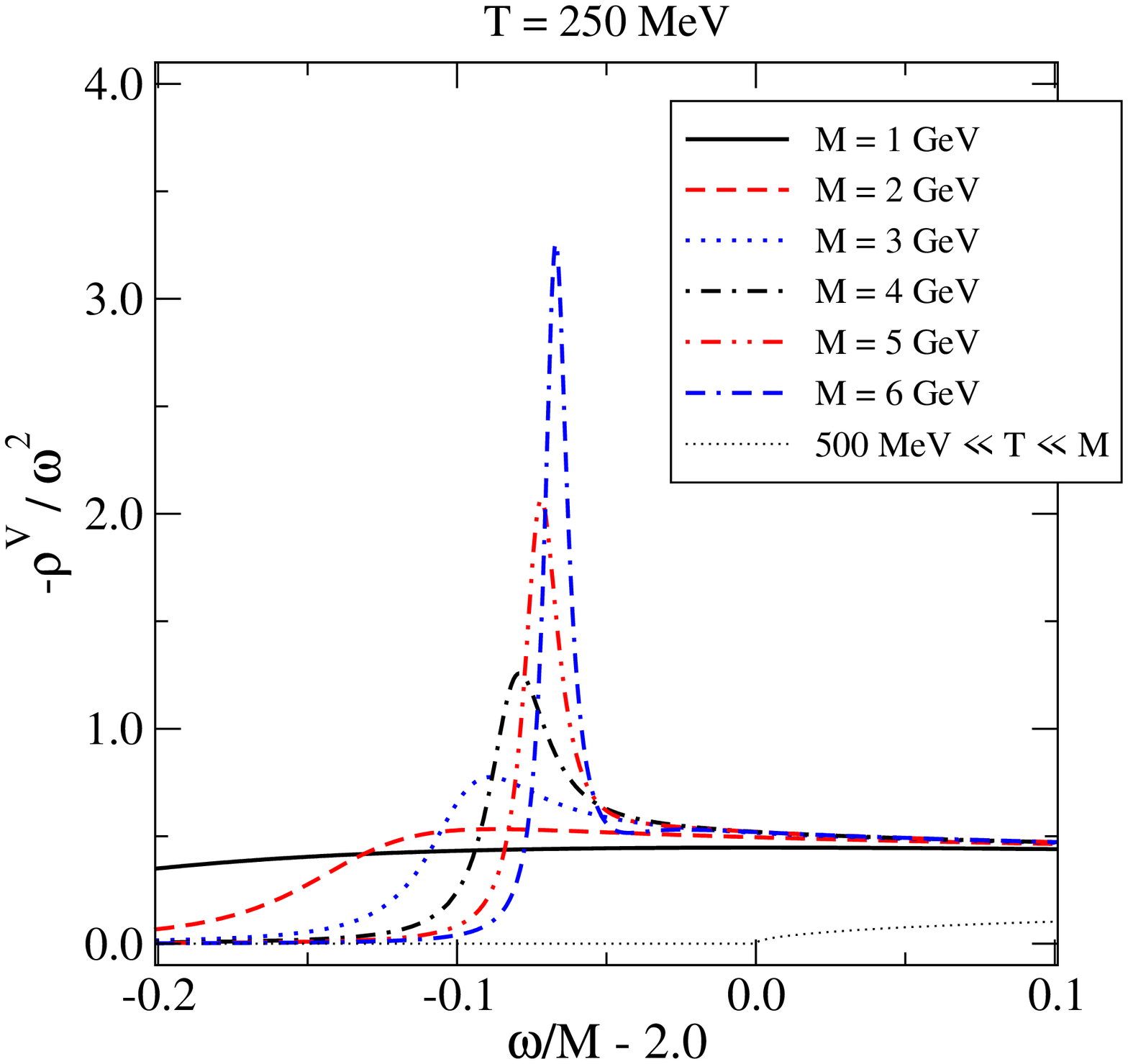}%
~~\epsfysize=5.0cm\epsfbox{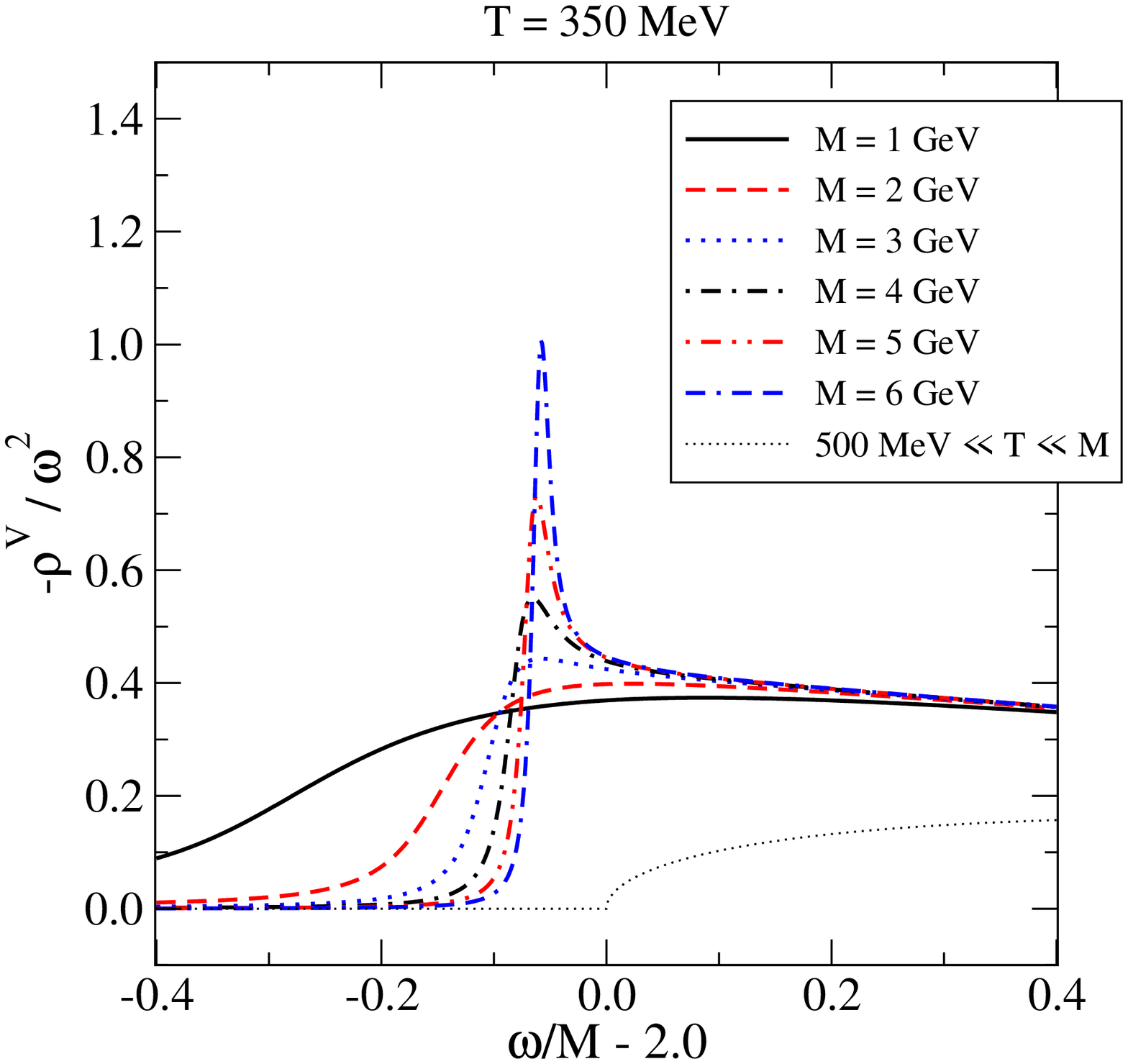}%
~~\epsfysize=5.0cm\epsfbox{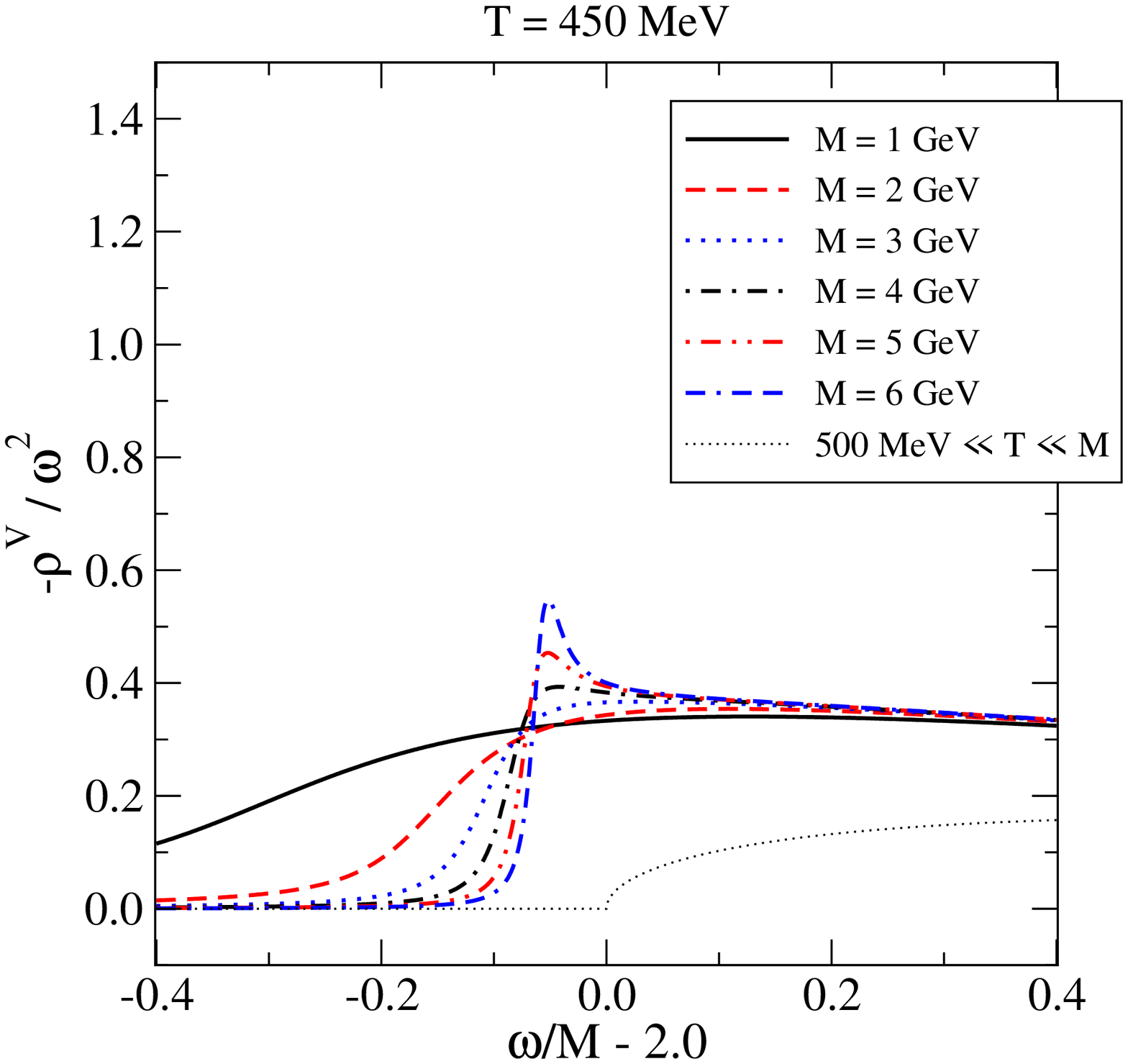}%
}

\vspace*{0.5cm}

%\vspace*{-4cm}

\caption[a]{The resummed perturbative vector channel
spectral function $\rho^V(\omega)$, in units of $\omega^2$, 
in the non-relativistic regime,
$(\omega - 2 M) / M \ll 1$, for $T = 250, 350, 450$~MeV (from left to right).
To the order considered, $M$ is the heavy quark pole mass. Note that
for better visibility, the axis ranges are different in the leftmost 
figure. 
}

\la{fig:rhoV_T}
\end{figure}
%%%%%%%%%%%%%%%%%%%%%%%%%%%%%%%%%%%%%%%%%%%%%%%%%%%%%%%%%%%%%%%%%%%%%%%%%%%

The results for $-\rho^V/\omega^2$ 
(\eq\nr{rho_res3} divided by $-\omega^2/M^2$) are shown 
in \figs\ref{fig:rhoV_M}, \ref{fig:rhoV_T}, and those for 
$\rho^S/\omega^2$ (\eq\nr{rhoS_full} divided by $\omega^2/M^2$)
in \figs\ref{fig:rhoS_M}, \ref{fig:rhoS_T}.
The results are given in a range of $\omega$ where relativistic
corrections, i.e.\ terms of higher order in a Taylor expansion
in $(\omega-2 M)/M$, are estimated to be at most at the 10\% level.
We show a scan of mass values, given that 
the inherent theoretical uncertainties of the charm and 
bottom pole masses are several hundred MeV 
(for a pedagogic discussion, see ref.~\cite{mb}), and 
that in lattice simulations there are further uncertainties, 
related to scale setting etc, which make it difficult 
to sit precisely at the physical point. 
As far as the other channels are concerned, 
we recall from \eqs\nr{CP}, \nr{CA0}, \nr{CACS_rel} that 
\be
 \rho^P \simeq -\fr13 \rho^V \;; \quad
 \rho^{A^0} \simeq -\fr13 \rho^V \;; \quad
 \rho^{\vec{A}} \simeq 2 \rho^S
 \;. 
\ee

%%%%%%%%%%%%%%%%%%%%%%%%% SUBSECTION %%%%%%%%%%%%%%%%%%%%%%%%%%%%%%%%%%
%
%\subsection{P-wave}

%%%%%%%%%%%%%%%%%%%%%%%%%%%%%%%%% FIGURE %%%%%%%%%%%%%%%%%%%%%%%%%%%%%%%%%
\begin{figure}[tb]

\centerline{%
\epsfysize=5.0cm\epsfbox{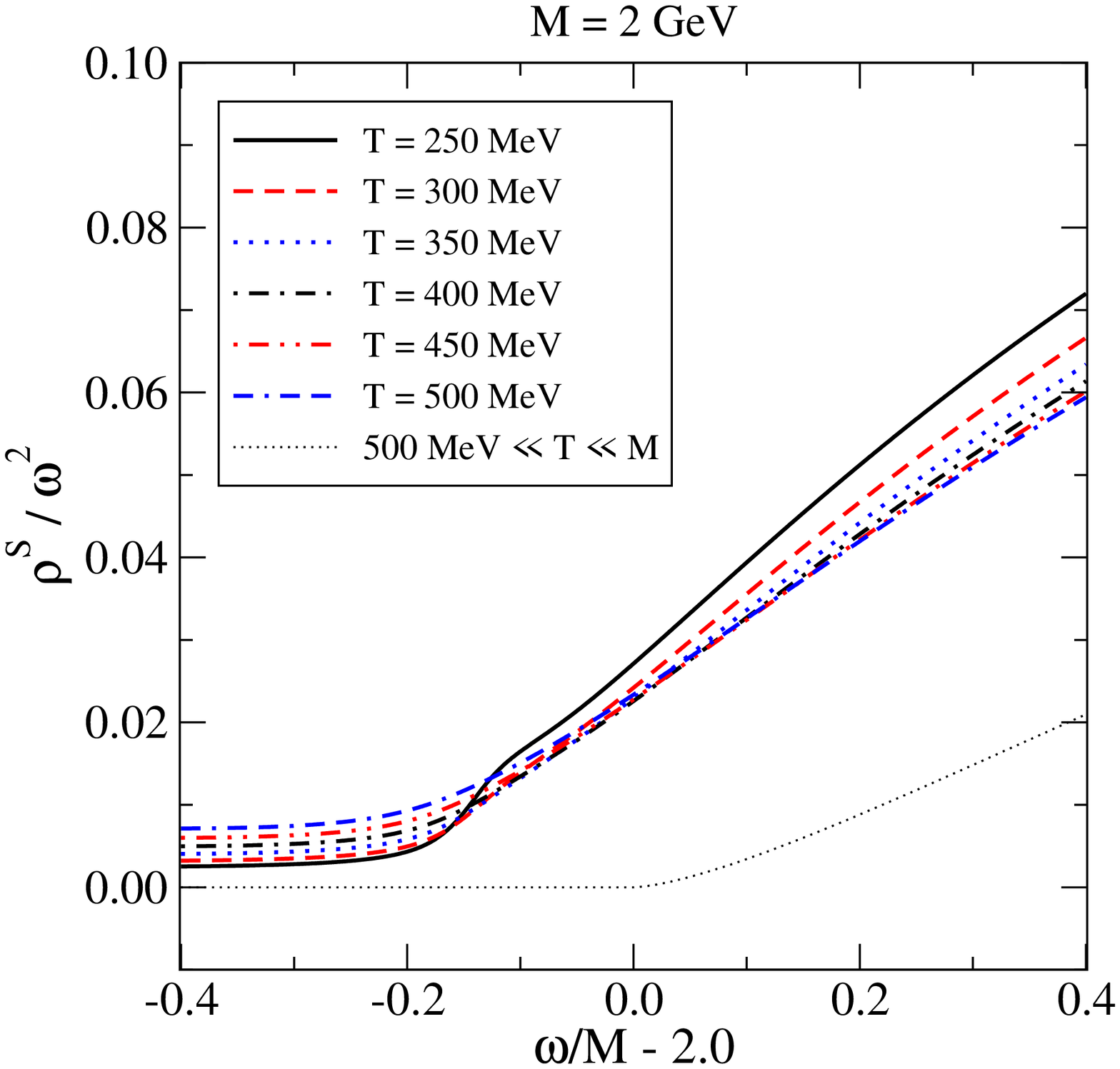}%
~~\epsfysize=5.0cm\epsfbox{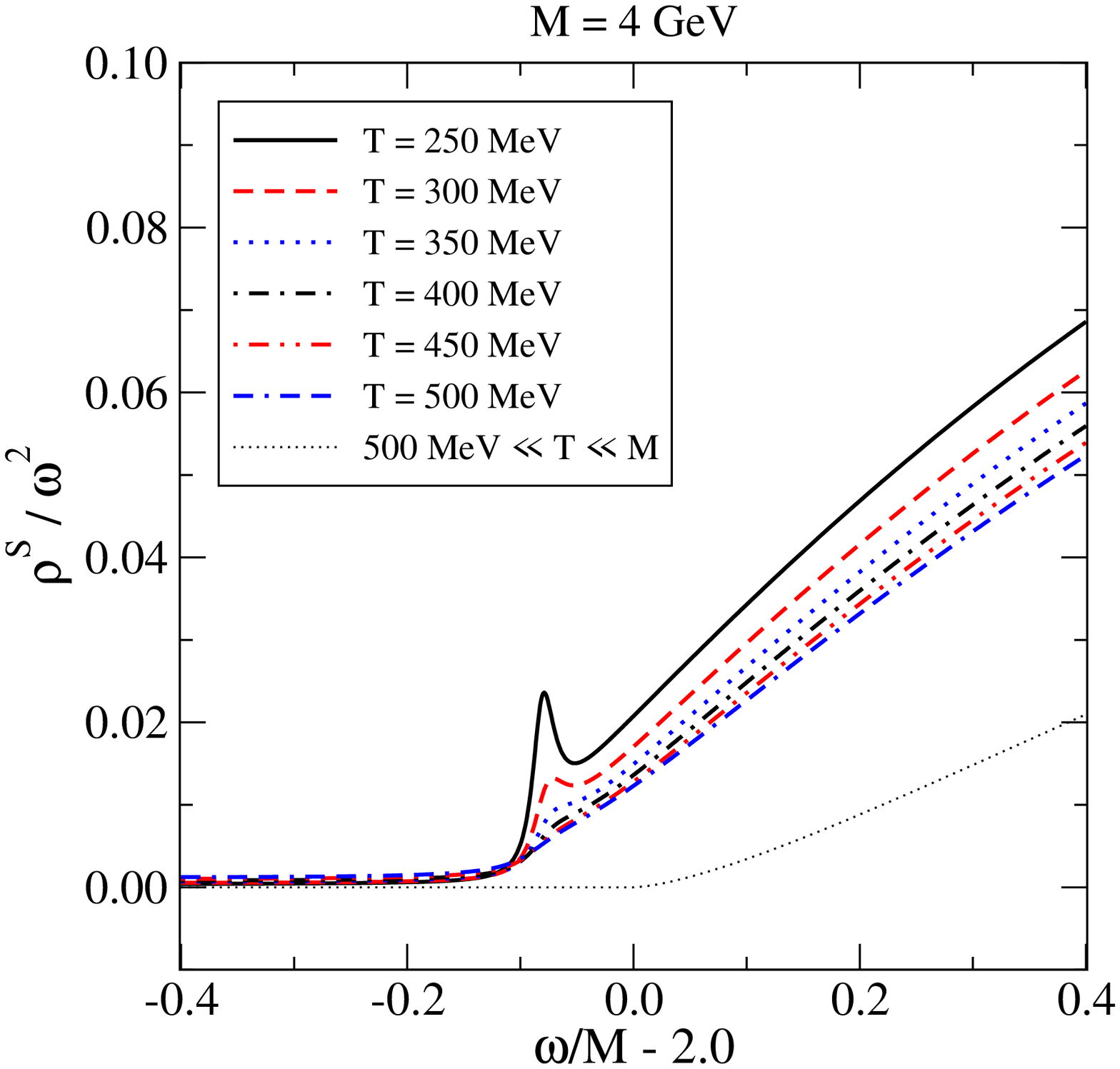}%
~~\epsfysize=5.0cm\epsfbox{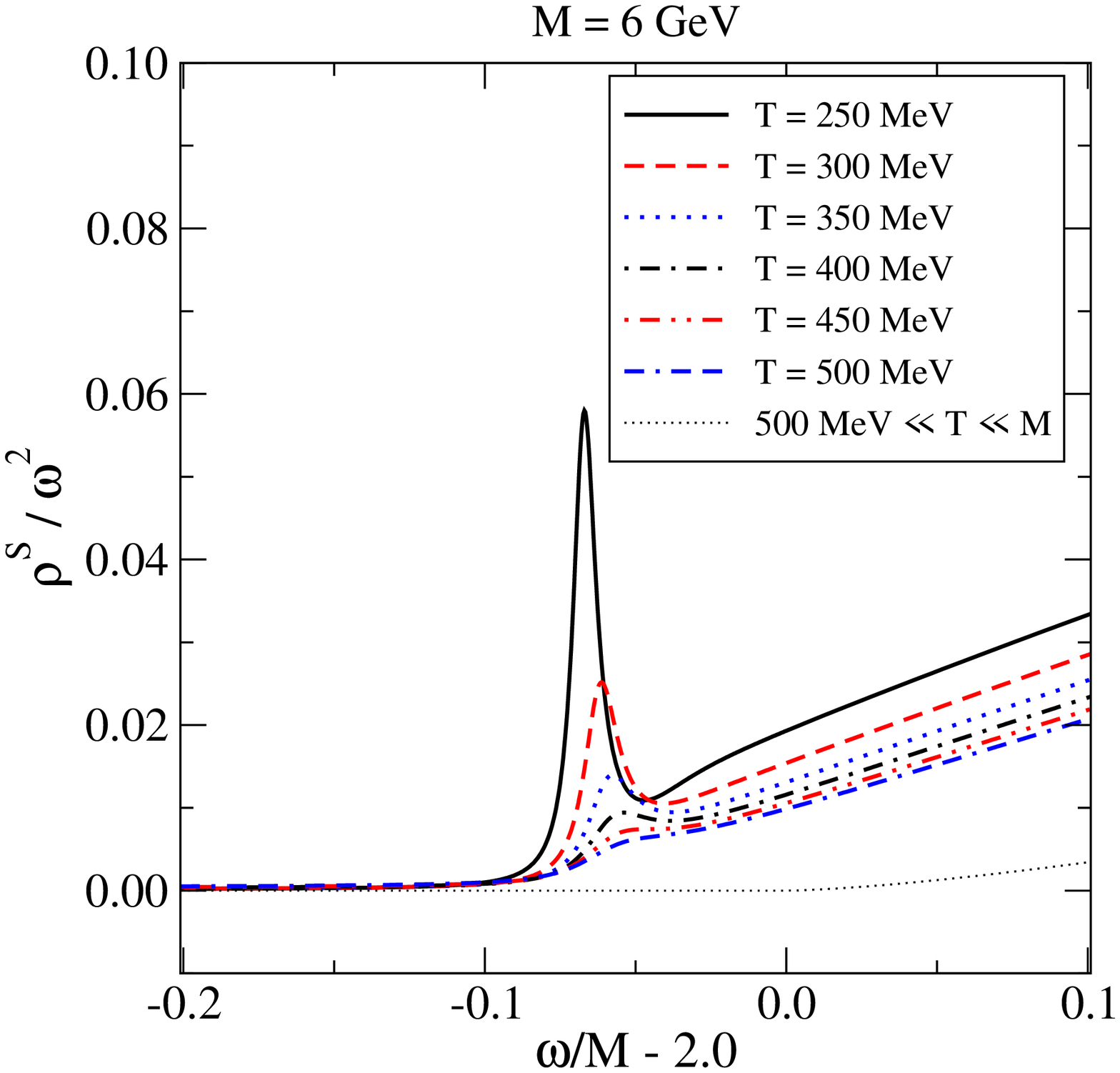}%
}

\vspace*{0.5cm}

%\vspace*{-4cm}

\caption[a]{The resummed perturbative scalar channel 
spectral function $\rho^S(\omega)$, in units of $\omega^2$, 
in the non-relativistic regime,
$(\omega - 2 M) / M \ll 1$, for $M = 2, 4, 6$~GeV (from left to right).
To the order considered, $M$ is the heavy quark pole mass. Note that
for better visibility, the range of $x$-axis is different in the rightmost 
figure. 
}

\la{fig:rhoS_M}
\end{figure}
%%%%%%%%%%%%%%%%%%%%%%%%%%%%%%%%%%%%%%%%%%%%%%%%%%%%%%%%%%%%%%%%%%%%%%%%%%%

%%%%%%%%%%%%%%%%%%%%%%%%%%%%%%%%% FIGURE %%%%%%%%%%%%%%%%%%%%%%%%%%%%%%%%%
\begin{figure}[tb]

\centerline{%
\epsfysize=5.0cm\epsfbox{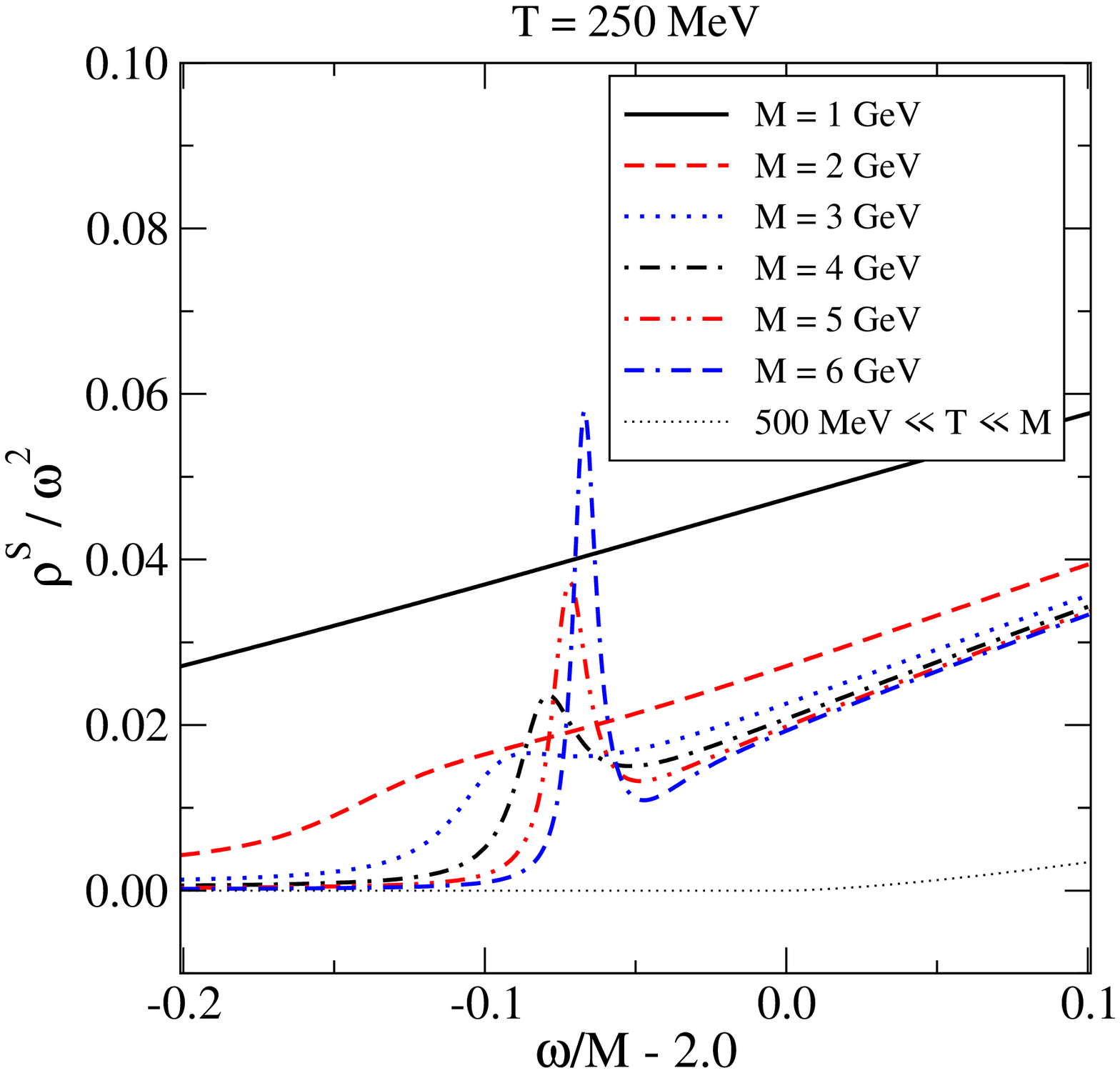}%
~~\epsfysize=5.0cm\epsfbox{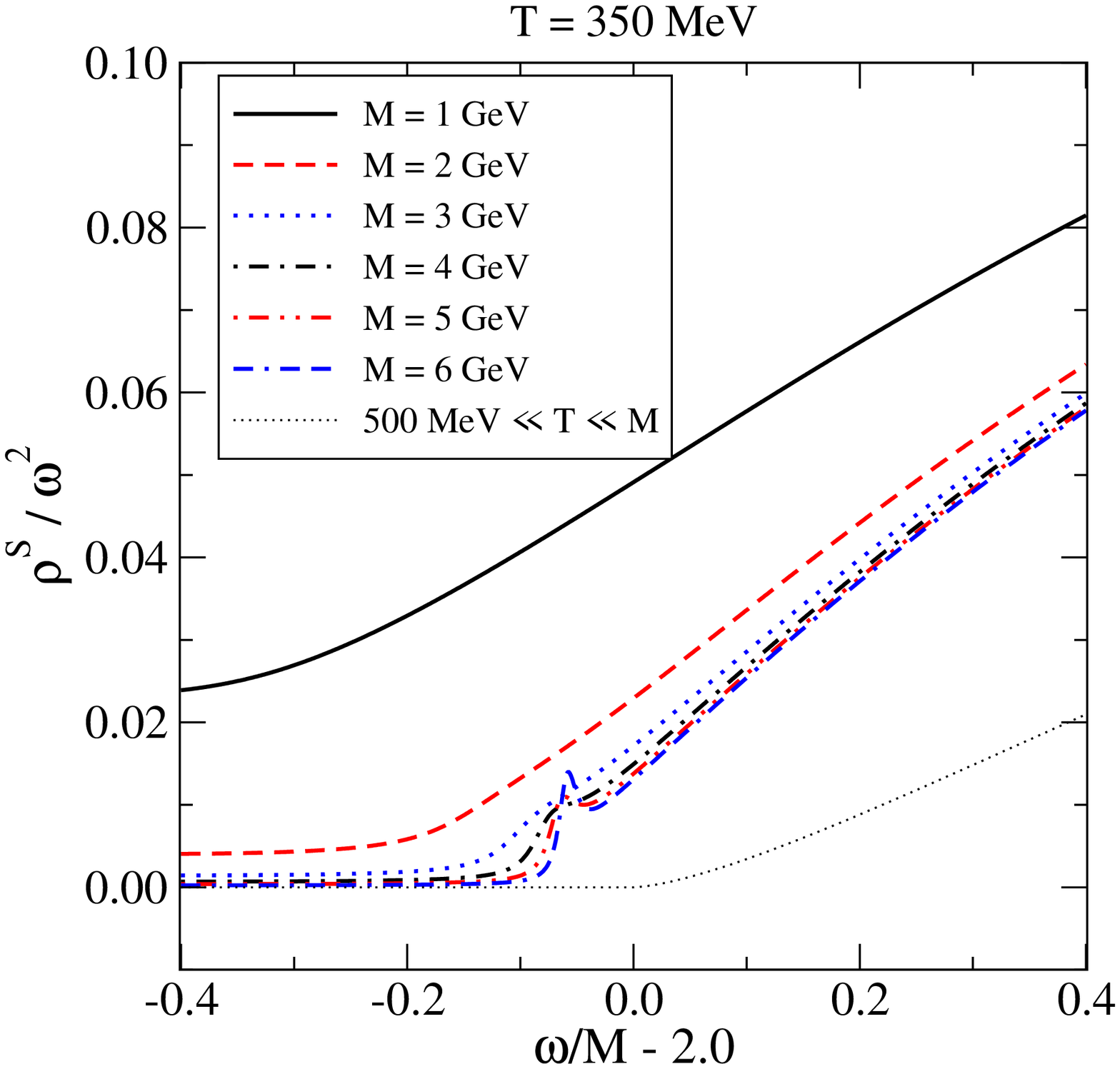}%
~~\epsfysize=5.0cm\epsfbox{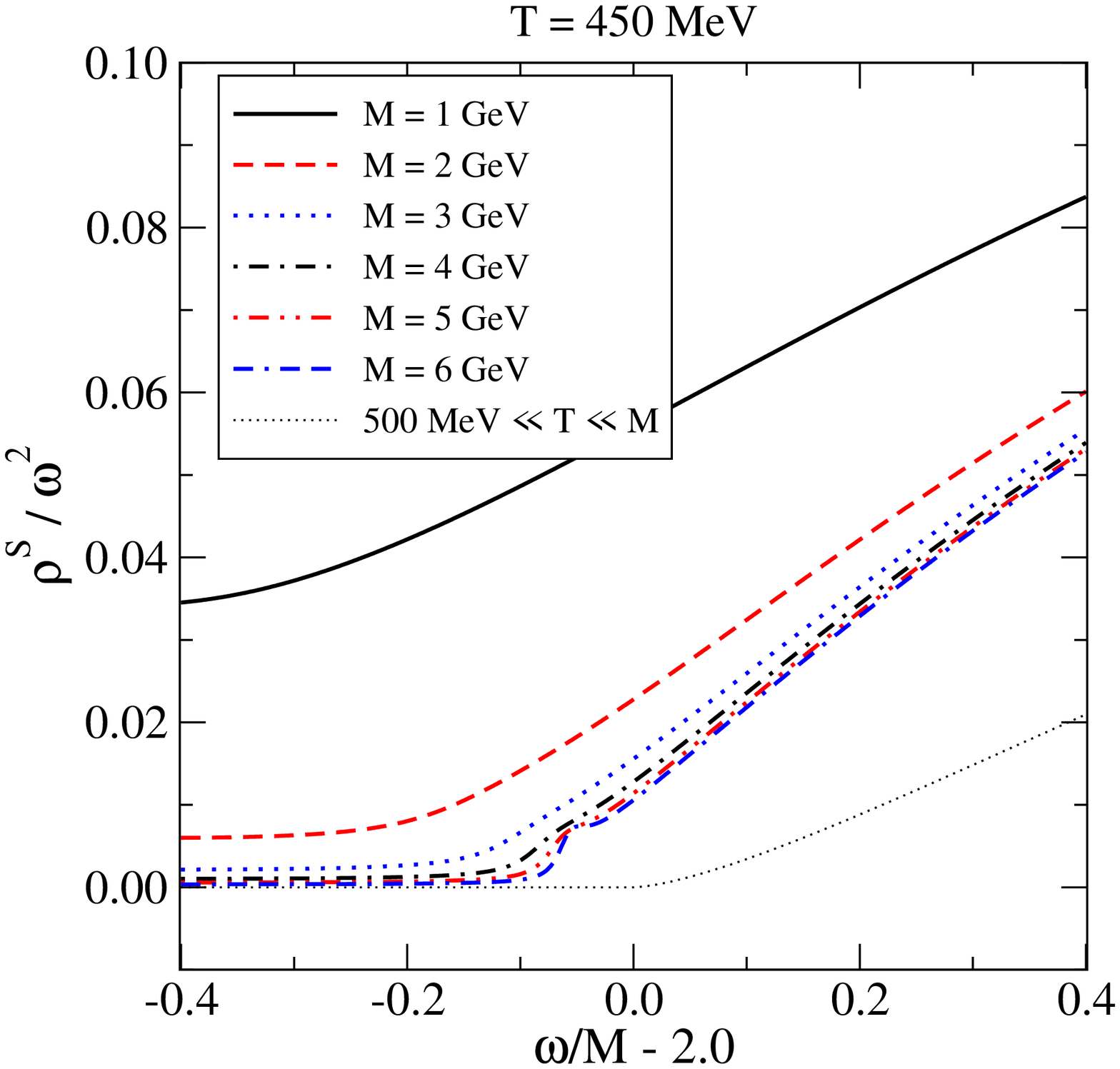}%
}

\vspace*{0.5cm}

%\vspace*{-4cm}

\caption[a]{The resummed perturbative scalar channel 
spectral function $\rho^S(\omega)$, in units of $\omega^2$, 
in the non-relativistic regime,
$(\omega - 2 M) / M \ll 1$, for $T = 250, 350, 450$~MeV (from left to right).
To the order considered, $M$ is the heavy quark pole mass. Note that
for better visibility, the range of $x$-axis is different in the leftmost 
figure. 
}

\la{fig:rhoS_T}
\end{figure}
%%%%%%%%%%%%%%%%%%%%%%%%%%%%%%%%%%%%%%%%%%%%%%%%%%%%%%%%%%%%%%%%%%%%%%%%%%%

%%%%%%%%%%%%%%%%%%%%%%%%% SUBSECTION %%%%%%%%%%%%%%%%%%%%%%%%%%%%%%%%%%
%
\subsection{Comparison with lattice}

As of today, lattice reconstructions of the spectral functions in 
various channels~\cite{lattold,latt1,latt2} 
suffer from significant uncertainties. 
Apart from the usual problems, it may be mentioned that the Compton
wavelength associated with the heavy quarks tends to be of the order
of the lattice spacing, so that we may expect even more 
significant discretization artifacts than in the usual quenched or 
2+1 light flavour simulations; and that the analytic continuation from 
Euclidean lattice data to the Minkowskian spectral function necessarily
involves model input, whose uncertainties are difficult to quantify. 
Nevertheless, it has been claimed that the latter 
types of uncertainties may be under reasonable control from a practical 
point of view~\cite{mem}. The most recent lattice results 
in this spirit can be found in refs.~\cite{latt1,latt2}.

It has become fashionable recently not to compare directly the 
spectral functions, but the Euclidean correlators for which direct
lattice data exists. Though this removes the uncertainties related
to the analytic continuation, it also comes with a heavy price: 
most of the structure in a Euclidean correlator is determined by 
values of $\omega$ far from the threshold, $\omega\ll 2M$
or $\omega \gg 2 M$,
so that the actual physics we are interested
in tends to be hidden in tiny effects somewhere in the middle 
of the Euclidean time interval.  For this reason, 
we do not consider Euclidean correlators
to be as interesting as the spectral functions, 
and touch only the latter in the following. 

Most of the lattice data exists for the charmonium case. 
The temperatures where the charmonium peak disappears from 
the spectral function are rather low, however; in fact they are 
in a regime where our analysis is probably not yet justified. 
Assuming the charmonium pole mass to be in the range 
$M\sim (1.5...2.0)$~GeV, we nevertheless observe from  
\fig\ref{fig:rhoV_M}(left) that at $T\approx 250$~MeV a certain
``enhancement'' can still be seen in the vector (and thus, 
in the pseudoscalar) channel. This then disappears at
higher temperatures. In contrast, in the scalar channel, 
\fig\ref{fig:rhoS_M}(left), there is practically no structure. 
These observations are certainly not in conflict 
with the lattice results of refs.~\cite{latt1,latt2}. 
Furthermore, we may note that the absolute magnitudes of 
$\rho_V$ and $\rho_S$ in \figs\ref{fig:rhoV_M}(left), 
\ref{fig:rhoS_M}(left) are qualitatively in a similar relation
to each other as the spectral functions measured on the lattice: 
the difference of about an order of magnitude 
is due to the $1/M^2$-suppression in the scalar 
case. At the same time, it needs to be kept in mind that in the 
scalar case the operators require renormalization, and that we 
have in any case not computed radiative corrections to the
absolute magnitudes of the spectral functions, so that the 
comparison cannot be taken too seriously. 

Data for the bottomonium case, where our predictions 
should be more reliable, can be found in ref.~\cite{latt1}.
There is again an inherent uncertainty of several hundred 
MeV in the bottom quark pole mass, but 
realistic values are presumable in the range $M\sim (4.5...5.0)$~GeV.
According to \figs\ref{fig:rhoV_M}, \ref{fig:rhoV_T} (middle to right), 
there is now 
a clear peak in the vector channel spectral function, up to 
a temperature of perhaps 500 MeV. In the scalar channel case, 
\figs\ref{fig:rhoS_M}, \ref{fig:rhoS_T} (middle to right), 
the structure is much 
less pronounced, but a tiny enhancement can be observed up to 
a temperature of about 400 MeV. These results are qualitatively
in better agreement with the lattice data in ref.~\cite{latt1} 
than the potential 
model results of ref.~\cite{mp3}, where no peak was found in the 
scalar channel case; as we have explained in \se\ref{se:rho}, 
the discrepancy can be traced back to a difference 
in the reconstruction of the spectral function from 
a Schr\"odinger equation. Nevertheless, in practice, it should again 
be stressed that systematic uncertainties of the lattice data are
certainly too large to make a quantitative comparison.

%%%%%%%%%%%%%%%%%%%%%%%%% SUBSECTION %%%%%%%%%%%%%%%%%%%%%%%%%%%%%%%%%%
%
\subsection{Dilepton rate}

%%%%%%%%%%%%%%%%%%%%%%%%%%%%%%%%% FIGURE %%%%%%%%%%%%%%%%%%%%%%%%%%%%%%%%%
\begin{figure}[ht]

\centerline{%
\epsfysize=7.5cm\epsfbox{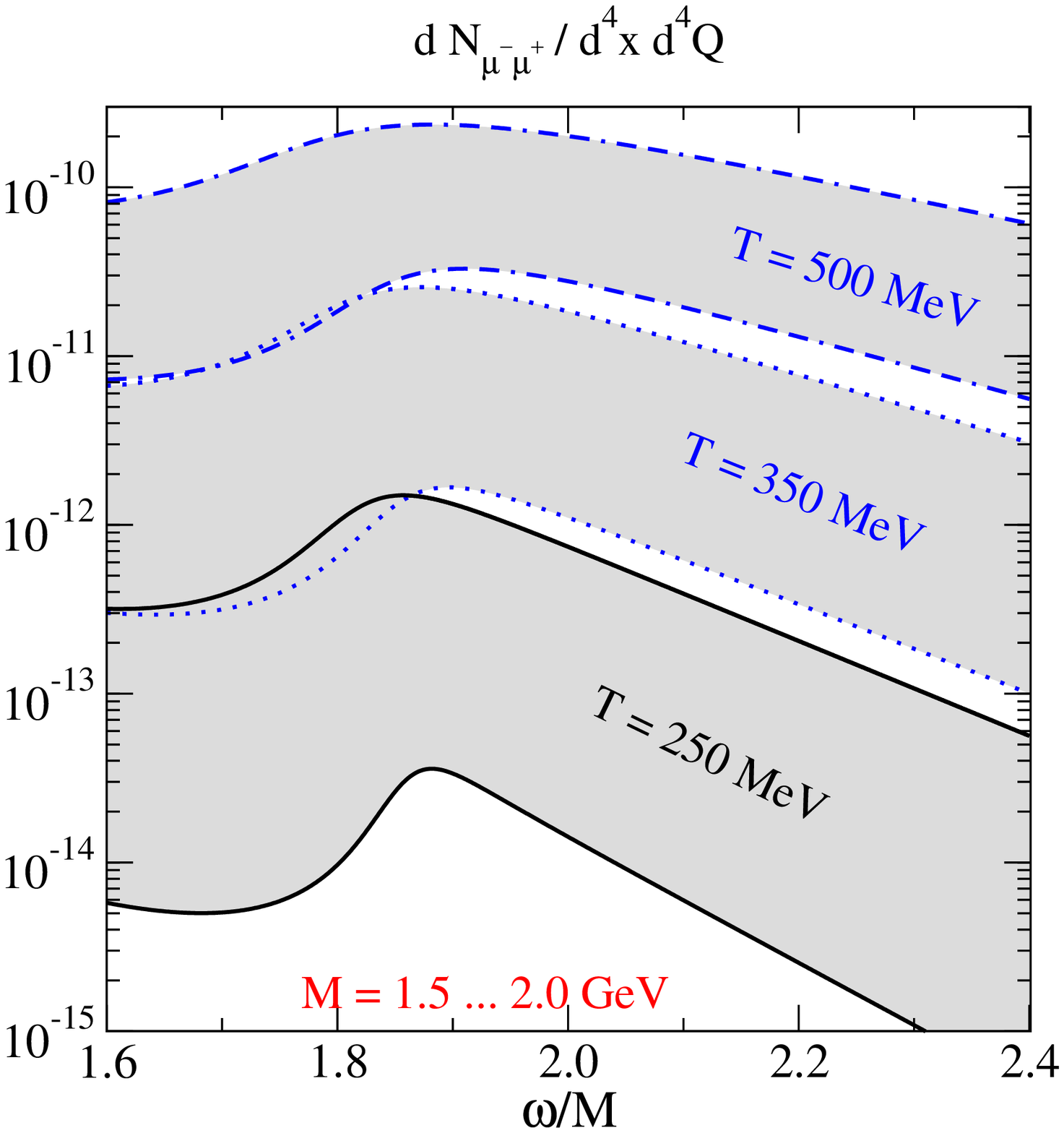}%
~~\epsfysize=7.5cm\epsfbox{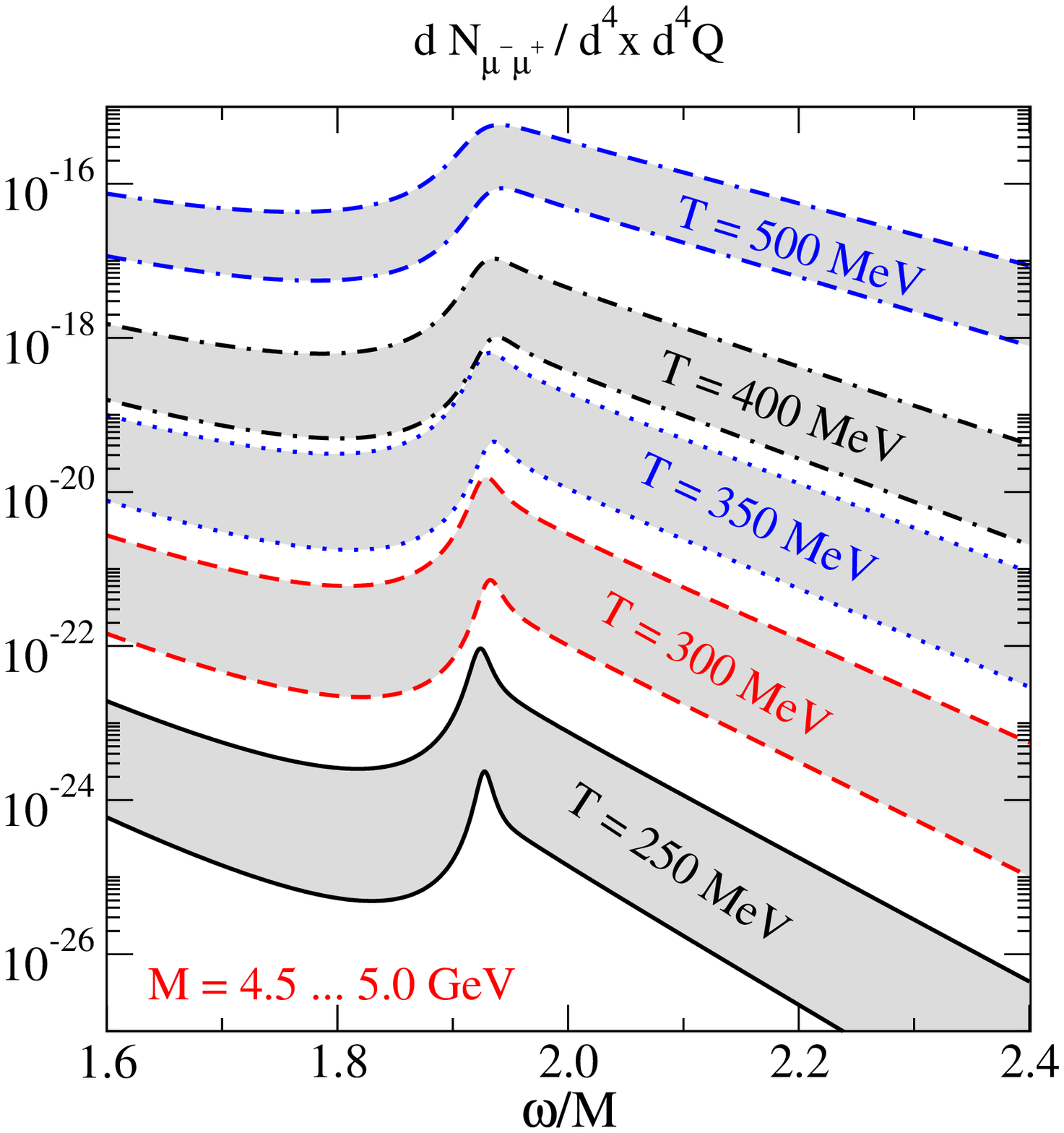}%
}

\vspace*{0.5cm}

%\vspace*{-4cm}

\caption[a]{The physical dilepton production rate, \eq\nr{dilepton}, 
from charmonium (left) and bottomonium (right), 
as a function of the energy, for various temperatures. 
The mass $M$ corresponds 
to the pole mass, and is subject to uncertainties of several
hundred MeV; we use the intervals 1.5...2.0 GeV and 4.5...5.0 GeV
to illustrate the magnitude of the corresponding error bands.
The low mass corresponds to the upper edge of each error band.}
\la{fig:rate}

\end{figure}
%%%%%%%%%%%%%%%%%%%%%%%%%%%%%%%%%%%%%%%%%%%%%%%%%%%%%%%%%%%%%%%%%%%%%%%%%%%

Apart from the spectral functions, it is interesting to plot
also the physical observable, the dilepton production rate 
given in \eq\nr{dilepton}. This is shown in \fig\ref{fig:rate}.
The significant difference with respect to the vector channel
spectral function is the existence of the Boltzmann factor 
(or, to be more precise, Bose-Einstein factor) 
in \eq\nr{dilepton}. Obviously, for a fixed frequency 
around the threshold, $\omega\sim 2 M$, the Boltzmann 
factor $\exp(-\omega/T)\sim \exp(-2 M/T)$ introduces a strong 
dependence of the dilepton rate on the temperature or, 
for a given temperature, on the mass. The exponential  
boosts the rate at high temperatures, and makes 
it decrease rapidly at large frequencies. Thereby the dilepton
rate shows a much stronger resonance-like behaviour than 
the spectral function, \fig\ref{fig:rhoV_M}. In particular, 
some kind of a peak structure remains visible in the dilepton rate 
in \fig\ref{fig:rate}
even at temperatures which are so high that there is only a smooth
step-like behaviour visible in the spectral function
in \fig\ref{fig:rhoV_M}. 

%%%%%%%%%%%%%%%%%%%%%%%%% SECTION %%%%%%%%%%%%%%%%%%%%%%%%%%%%%%%%%%%%%
%
\section{Physical picture of heavy quarkonium in a thermal plasma}

Conceptually, the most important difference between our analysis 
and traditional potential models~\cite{models,mp3} is the existence
of an imaginary part in the static potential, \eq\nr{expl}.
Physically, the imaginary part
implies that quarkonium at high temperatures should not 
be thought of as a stationary state. Rather, the norm of its 
wave function decays exponentially with (Minkowski) time.
This is due to the fact that, apart from experiencing Debye screening, 
there is also a finite probability for the off-shell gluons
binding the two quarks to disappear, due to Landau damping, 
i.e.\ inelastic scatterings with hard particles in the plasma.
Once $T\sim gM$,  
the imaginary part is in fact parametrically larger
than the binding energy (cf.\ \se\ref{se:power}).
At the same time, for low enough temperatures, $T\sim g^2 M$, 
the imaginary part plays a subdominant role (cf.\ \se\ref{se:power}). 

It may be useful to remark that if, 
on the contrary, one goes to a Euclidean lattice, then a non-zero 
wave function {\em can} be defined at any finite value 
of the ``imaginary time'' 
coordinate $\tau$, $0 < \tau < \beta$. 
Introducing also gauge-fixing, such wave functions have 
been measured with Monte Carlo simulations in ref.~\cite{wf}
(for a recent review, see ref.~\cite{wfrev}).
With regard to the discussion above, 
the physical significance of such wave functions 
for Minkowski-time observables is not obvious; 
hence we do not discuss them here.

%%%%%%%%%%%%%%%%%%%%%%%%% SECTION %%%%%%%%%%%%%%%%%%%%%%%%%%%%%%%%%%%%%
%
\section{Conclusions}

The purpose of this paper has been to experiment, 
as generally as possible,  with the resummed
perturbative framework that was introduced in refs.~\cite{static,og2}, 
in order to offer one more handle on the properties 
of heavy quarkonium in hot QCD, thus supplementing the traditional 
approaches based on potential models and on lattice QCD. 

The key ingredient of our approach is a careful definition of 
a finite-temperature real-time static potential that can be 
inserted into a Schr\"odinger equation obeyed by certain heavy
quarkonium Green's functions. The potential in question, denoted 
by $\lim_{t\to\infty}V_{>}(t,r)$, has both a real and an imaginary 
part (cf.\ \eq\nr{expl}). 
An important conceptual consequence from the existence of an
imaginary part is that heavy quarkonium should
not be thought of as a stationary state at high temperatures, 
but as a short-lived transient, 
with the quark and antiquark binding together 
only for a brief moment before unattaching again.  

On the more technical level we have noted that, 
in terms of \eq\nr{omega_rep}, 
the vector channel spectral function gets
a contribution only from the S-wave, $l=0$, while the scalar
channel spectral function gets a contribution both from the S-wave 
and P-wave, $l=0,1$. Here we differ from the potential model
analysis in ref.~\cite{mp3} where, as far as we can see, 
only $l=1$ was considered for the scalar channel. 
The reason for the difference is 
discussed at the end of \se\ref{se:rho}. The difference is 
significant, since the S-wave contribution introduces a small
reasonance peak to the scalar channel spectral function as well. 

The phenomenological pattern
we find for the spectral functions within this framework 
is not too different from indications from lattice QCD: scalar
channel charmonium displays practically no reasonance peak above 
a temperature of 200 MeV; vector channel charmonium has some peak-like
structure up to a temperature of about 300 MeV; scalar channel 
bottomonium is again weakly bound but does show a small enhancement 
up to a temperature of about 400 MeV; vector channel bottomonium 
can support a resonance peak up to a temperature of about 500 MeV. 
(Because of unknown higher order corrections, 
these numbers are subject to uncertainties 
of several tens of MeV.)

At the same time,  we stress that in the physical dilepton rate, 
\fig\ref{fig:rate}, the quarkonium peak always  becomes
{\em more pronounced} with increasing temperature, irrespective of 
the disappearance of the resonance structure from the spectral function.  
This boost is due to an interplay of the free quark continuum in the 
spectral function, and the Boltzmann factor $\exp(-\omega/T)$.

There are a few directions in which our work could be extended,
in order to go beyond a purely perturbative approach. 
In particular, the imaginary part of the real-time static potential
has been measured with classical lattice gauge theory simulations 
in ref.~\cite{imV}, and could thus to some extent be used in 
a non-perturbative setting. Hopefully, the real part of our
static potential could also be related to quantities that 
are measurable with lattice Monte Carlo methods, thereby allowing us 
to probe more reliably the phenomenologically  interesting
temperature regime around a few hundred MeV.

%%%%%%%%%%%%%%%%%%%%%%%%% SECTION %%%%%%%%%%%%%%%%%%%%%%%%%%%%%%%%%%%%%
%
\section*{Acknowledgements}

M.L.\ thanks T.~Hatsuda and S.~Kim for useful suggestions, 
and the  Isaac Newton Institute for Mathematical Sciences, 
where part of this work was carried out, for hospitality.
This work was partially supported by the BMBF project
{\em Hot Nuclear Matter from Heavy Ion Collisions 
     and its Understanding from QCD}.

%\newpage

%-------------------------------------------------------------------

\appendix
\renewcommand{\thesection}{Appendix~\Alph{section}}
\renewcommand{\thesubsection}{\Alph{section}.\arabic{subsection}}
\renewcommand{\theequation}{\Alph{section}.\arabic{equation}}

%%%%%%%%%%%%%%%%%%%%%%%%% SECTION %%%%%%%%%%%%%%%%%%%%%%%%%%%%%%%%%%%%%
%
\section{Numerical method for finding the spectral functions}

In this appendix we provide details 
concerning the numerical method that we have 
used for determining the vector and scalar channel spectral functions. 
The basic approach is from ref.~\cite{ps}, where it was applied
for the vector channel at zero temperature; the method was extended
to the scalar channel case in ref.~\cite{mp3}. Our presentation is
rather close to that in ref.~\cite{mp3}, but we choose to spell out the 
details anew due to the fact that, as already mentioned in \se\ref{se:rho}, 
we find one additional term in the scalar channel case. Furthermore, 
the existence of an imaginary part in our static potential simplifies 
certain points of the analysis. We should point out that the method 
presented here appears to be numerically superior to that introduced for 
the vector channel in ref.~\cite{og2}. 

\subsection{Vector channel}
\la{ss:S}

We proceed with the evaluation of \eq\nr{rho_res}.
Given the ansatz in \eq\nr{ansatz}, 
it remains to determine $A, g_{<}^{l}, g_{>}^{l}$,
and then to extrapolate $r,r'\to 0$. We thus need to know, 
in particular, the asymptotic behaviours of the functions
$g_{<}^{l}, g_{>}^{l}$ near the origin. 
Let $g^l_\rmi{r}$ and $g^l_\rmi{i}$ be the solutions regular 
and irregular around the origin, respectively: 
\ba
 g^l_\rmi{r} & = & r^{l+1} \sum_{n=0}^{\infty} a_n r^n \approx a_0 r^{l+1}
 \;, \la{glr} \\ 
 g^l_\rmi{i} & = & g^l_\rmi{r} (r) 
 \int_{\delta}^r \! {\rm d} r' \, \frac{1}{[g^l_\rmi{r}(r')]^2}
         \approx -\frac{1}{a_0} \frac{r^{-l}}{2l+1}
 \;. \la{gli}
\ea
We may then choose 
\ba
 g^l_{<}(r) & = & g^l_\rmi{r}(r) 
 \;, \la{glr2} \\ 
 g^l_{>}(r) & = & g^l_\rmi{i}(r) + B^l g^l_\rmi{r}(r)
 \;, \la{gli2}
\ea
where the coefficient $B^l$ is defined such as to guarantee 
the regularity of $g^l_{>}(r)$ at infinity, 
\be
 B^l = -\lim_{r\to\infty} \frac{g^l_\rmi{i}(r)}{g^l_\rmi{r}(r)}
     = - \int_{\delta}^{\infty} \! {\rm d} r' \, \frac{1}{[g^l_\rmi{r}(r')]^2}
 \;. \la{Bl}
\ee
Combining \eqs\nr{gli}, \nr{gli2}, \nr{Bl}, we can write
\be
 g^l_{>}(r) = - g^l_\rmi{r}(r)
 \int_r^\infty \! {\rm d} r' \, \frac{1}{[g^l_\rmi{r}(r')]^2}
 \;. \la{gli3}
\ee

Let us next compute the coefficient $A$ in \eq\nr{ansatz}.
Integrating both sides of \eq\nr{Seq_omega_l} with 
$ 
 \int_{r'-0^+}^{r'+0^+} \! {\rm d}r \, (...)
$,
yields
\be
 A = \frac{6 \Nc M}
          {g_{>}^{l}(r') {\rm d} g_{<}^{l}(r') / {\rm d} r' - 
           g_{<}^{l}(r') {\rm d} g_{>}^{l}(r') / {\rm d} r'}
 \;. \la{A}
\ee
Involving a Wronskian, this expression is independent of the 
position $r'$ at which it is evaluated, so we can do 
this at small $r'$. Then we can use the asymptotic 
forms from \eqs\nr{glr}, \nr{gli}, to find that
\be
 A = -6 \Nc M \;. \la{AA}
\ee
Note that this expression is independent of $l$.

We finally take the limit $r,r'\to 0$, while keeping 
$r < r'$, so that $r_{<} \equiv r$, $r_{>} \equiv r'$.
Inserting \eqs\nr{ansatz}, \nr{glr}, and \nr{AA} 
into \eq\nr{rho_res},  yields
\ba
 \rho^V(\omega') & = &  
 \frac{6 \Nc M }{4\pi}
 \lim_{r,r'\to 0} \frac{1}{rr'}
 \im \Bigl[ g_{>}^0 (r') g_{<}^0 (r) \Bigr] 
 \nn & = &  
 - \frac{6 \Nc M a_0}{4\pi}
 \lim_{r'\to 0} 
 \im \biggl\{ \frac{g^0_\rmi{r} (r')}{r'} 
 \int_{r'}^\infty \! {\rm d}r'' \, \frac{1}{[g^0_\rmi{r}(r'')]^2}
 \biggr\} 
 \;, \la{rho_res2}
\ea
where we assumed $a_0$ to be real.

Let us now analyse the origin of the imaginary part in \eq\nr{rho_res2}.
It will be convenient to express the $r$-dependence in terms of 
the dimensionless variable $\varrho \equiv r \alpha M$, 
where $\alpha \equiv g^2 C_F/4\pi$. In these units, 
the homogeneous Schr\"odinger equation (\eq\nr{Seq_omega_l_hom}) reads
\be
 \biggl[
   \frac{\partial^2}{\partial\varrho^2}
  - \frac{l(l+1)}{\varrho^2} + \frac{1}{\varrho} + \rmO(1) 
 \biggr] g^l_\rmi{r}(\varrho) = 0 
 \;, 
\ee
implying
\be
 g^l_\rmi{r}(\varrho) = \varrho^{l+1} - \frac{1}{2(l+1)} \varrho^{l+2} + ...
 \;. \la{glr_expansion}
\ee
At some order the solution also develops an imaginary part; 
let us write an ansatz
\be
 g^l_\rmi{r}(\varrho) = \varrho^{l+1} - \frac{1}{2(l+1)} \varrho^{l+2} + \ldots
 + i \gamma_1 \varrho^x 
 \;, \quad \gamma_1 \in \RR \;. \la{small_rho}
\ee
The imaginary part in \eq\nr{Seq_omega_l} 
behaves as $\sim i \gamma_2 \varrho^2$
at small $\varrho$. Inserting into the Schr\"odinger equation, 
we get for the leading imaginary term
\be
 i \gamma_1 \varrho^{x-2}[x(x-1) - l(l+1)] 
+ i \gamma_2 \varrho^2 \cdot \varrho^{l+1} = 0
 \;, \la{impart_expansion} 
\ee
implying $x= l+5$. 

Returning to \eq\nr{rho_res2}, there are in principle 
two possibilities for the origin of the imaginary part. 
However, according to \eqs\nr{glr_expansion}, \nr{impart_expansion}, 
$ 
 \lim_{r'\to 0} \im[g_\rmi{r}^0/r']\int_{r'}^\infty {\rm d}r'' 
 \re\{ 1/[g_\rmi{r}^0(r'')]^2 \} \sim 
 \lim_{r'\to 0} (r')^4 /(r') = 0
$.
Therefore the imaginary part can only arise from 
$
 \im \{ 1/[g_\rmi{r}^0(r'')]^2 \}
$. 
Inserting the asymptotic form of 
$
 \re[g_\rmi{r}^0/r']
$
from \eq\nr{glr_expansion}; using the variable $\varrho$; 
and noting that this  corresponds to the choice $a_0 = \alpha M$,
we then obtain \eq\nr{rho_res3}. 

It is useful to crosscheck that \eq\nr{rho_res3} produces 
the correct result in the free limit. In the free case there is 
no $i |\im V_{>}(r)|$ in \eq\nr{Seq_omega_l}, 
and a factor $i\epsilon \equiv i0^+$ needs to be 
inserted instead, to pick up the correct (retarded) solution. 
In dimensionless units, the homogeneous equation then becomes
\be
 \biggl[ 
 \frac{\partial^2}{\partial \varrho^2}
 + \frac{\hat\omega'}{\alpha^2} + i \epsilon
 \biggr] g^0_\rmi{r} (\varrho)  
 = 0 
 \;,
\ee
where $\hat\omega' \equiv \omega'/M$. We denote 
\be
 k \equiv \sqrt{\frac{\hat\omega'}{\alpha^2}+i \epsilon}
 \;. \la{k_def}
\ee
The solution with the correct behaviour around the origin
(with $a_0 = \alpha M$) reads
\be
 g^0_\rmi{r} (\varrho) =  
 \frac{1}{k}
 \sin( k\varrho ) \;.
\ee
We can write 
\be
  \frac{1}{[g^0_\rmi{r}(\varrho')]^2} 
  = 
 - k 
 \frac{{\rm d}}{{\rm d}\varrho'} \biggl\{
 \frac{\cos\bigl( k \varrho' \bigr)}
      {\sin\bigl( k \varrho' \bigr)} \biggr\}
 \;.
\ee 
The integral in \eq\nr{rho_res3} can now be carried out; the substitution
at the upper end gives a contribution from the exponentially growing
terms $\exp(-i k \varrho' )$, present both in 
the cosine and in the sine. Their ratio gives $-i$, and the total 
is then 
\be
 \frac{\rho^V(\omega')}{M^2} = 
 - \frac{6 \Nc \alpha }{4\pi}
 \im\biggl\{
    \biggl({\frac{\hat\omega'}{\alpha^2}+i \epsilon}\biggr)^{\fr12} i 
  \biggr\}
  = -\frac{3\Nc}{2\pi}\, \theta(\omega') \, (\hat \omega')^{1/2}
 \;.
\ee  
This indeed agrees with \eq\nr{tree_S}.

\subsection{Scalar channel}

In the scalar channel case, the equations to be solved 
are \nr{rhoS_Psi}, \nr{omega_rep}, \nr{Seq_omega_l}; 
the ansatz for the solution is in \eq\nr{ansatz}, 
with $A$ given by \eq\nr{AA}. 

Let us first work out the contribution from the mode $l=0$ (S-wave). 
According to \eqs\nr{omega_rep}, \nr{ansatz}, \nr{glr2}, \nr{gli3}, 
the relevant term of $\tilde\Psi$, denoted by $\delta_o\tilde\Psi$, is 
\be
 \delta_0 \tilde \Psi(\omega';\vec{r,r'})
 = - \frac{1}{4\pi r r'} A\, 
 g_\rmi{r}^0(r) g_\rmi{r}^0(r')
 \int_{r'}^\infty \! {\rm d}r'' \, \frac{1}{[g^0_\rmi{r}(r'')]^2}
 \;.
\ee
Inserting into \eq\nr{rhoS_Psi}, making use of \eq\nr{AA},  
and going over into the dimensionless variable $\varrho$,  we get
\be
 \frac{\delta_0 \rho^S(\omega')}{M^2}
 = \frac{2\Nc \alpha^3}{4\pi}
 \lim_{\varrho,\varrho'\to 0}
 \im
 \biggl\{ 
   \frac{{\rm d}}{{\rm d}\varrho}
   \biggl( \frac{g_\rmi{r}^0(\varrho)}{\varrho} 
   \biggr)
   \frac{{\rm d}}{{\rm d}\varrho'}
   \biggl( 
    \frac{g_\rmi{r}^0(\varrho')}{\varrho'} 
    \int_{\varrho'}^{\infty}
    \! {\rm d}\varrho'' \, \frac{1}{[g_\rmi{r}^0(\varrho'')]^2} 
   \biggr)
 \biggr\} 
 \;. \la{new_1}
\ee
According to \eq\nr{glr_expansion}, the first term inside the curly
brackets is 
$
 \lim_{\varrho\to 0} {\rm d}_\varrho ( g_\rmi{r}^0 / \varrho) 
 = -1/2
$, so that we get
\be
 \frac{\delta_0 \rho^S(\omega')}{M^2}
 = - \frac{\Nc \alpha^3}{4\pi}
 \lim_{\varrho'\to 0}
 \im
 \biggl\{ 
   \frac{{\rm d}}{{\rm d}\varrho'}
   \biggl( 
    \frac{g_\rmi{r}^0(\varrho')}{\varrho'} 
    \int_{\varrho'}^{\infty}
    \! {\rm d}\varrho'' \, \frac{1}{[g_\rmi{r}^0(\varrho'')]^2} 
   \biggr)
 \biggr\} 
 \;. \la{new_2}
\ee

In principle there are again two possible origins for the imaginary 
part. However, as we saw in the vector channel case, 
$ 
 \im[g_\rmi{r}^0/\varrho']
 \int_{\varrho'}^\infty {\rm d}\varrho'' 
 \re\{ 1/[g_\rmi{r}^0(\varrho'')]^2 \} \sim 
 (\varrho')^3
$, 
so that a non-zero contribution can only arise from 
$
 \im \{ 1/[g_\rmi{r}^0(\varrho'')]^2 \}
$. 
Furthermore, the derivative can only act on the combination
multiplying the integral, since
\be
 \frac{\re [g_\rmi{r}^0(\varrho')]}{\varrho'}
  \im \biggl\{ \frac{1}{[g_\rmi{r}^0(\varrho')]^2} \biggr\} 
 \approx 
 \im \biggl\{
 \frac{1}{[g_\rmi{r}^0(\varrho')]^2} 
 \biggr\}
 \approx  \im \biggl\{
 \frac{1}{(\varrho')^2 + 2 i \gamma_1 (\varrho')^6} 
 \biggr\}
 \approx
 -2 \gamma_1 (\varrho')^2
 \;.  \la{ests_1} 
\ee
Making use of 
$
 \lim_{\varrho'\to 0} {\rm d}_{\varrho'} ( g_\rmi{r}^0 / \varrho') 
 = -1/2
$, 
the S-wave contribution to the scalar spectral function thus becomes
\be
 \frac{\delta_0 \rho^S(\omega')}{M^2}
 = \frac{\Nc \alpha^3}{8\pi}
 \lim_{\delta\to 0} \left.
    \int_{\delta}^{\infty} \! {\rm d} \varrho \,
    \im \biggl\{ \frac{1}{[g_\rmi{r}^0(\varrho)]^2}  \biggr\}
    \right|_{g^0_\rmi{r}(\varrho) = \varrho - \varrho^2/2 + ...}
 \;. \la{new_3}
\ee 
In other words, comparing with \eq\nr{rho_res3}, 
$\delta_0 \rho^S(\omega') = - \alpha^2 \rho^V(\omega')/12$;
the factor $\alpha^2$ is a manifestation of the suppression
$\sim \nabla_\vec{r}^2/M^2$ apparent in \eq\nr{CVCS_rel}, 
combined with the parametric order of magnitude of 
$\nabla_\vec{r}/M$ from \eq\nr{param_magn}.

Consider then the contribution from the mode $l=1$ (P-wave).
The relevant term from \eq\nr{ansatz}, 
denoted by $\delta_1\tilde\Psi$, is
\be
 \delta_1 \tilde \Psi(\omega';\vec{r,r'})
 = 
 A \frac{g_{<}^{1}(r)}{r} \frac{g_{>}^{1}(r')}{r'}
 \sum_{m=-1}^{1}
 Y_{1m}(\theta,\phi) Y_{1m}^*(\theta',\phi')
 \;.
\ee
Hence we will need
\be
 Y_{10}(\theta,\phi) = \sqrt{\frac{3}{4\pi}} \cos\theta
 \;, \quad
 Y_{1\pm 1}(\theta,\phi) = \mp 
 \sqrt{\frac{3}{8\pi}} \sin\theta e^{\pm i\phi}
 \;. \la{Ylm}
\ee
In order to take the derivatives in \eq\nr{rhoS_Psi}, 
we stay with radial coordinates, so that 
\be
 \nabla_\vec{r} = 
 \vec{e}_r \frac{\partial}{\partial r} + 
 \vec{e}_\theta \frac{1}{r} \frac{\partial}{\partial\theta} + 
 \vec{e}_\phi \frac{1}{r\sin\theta} \frac{\partial}{\partial\phi}
 \;. \la{nabla_rad}
\ee
Moreover, we choose again $r<r'$, so that 
$r_{<} \equiv r$, $r_{>} \equiv r'$.
We will set $\Omega' = \Omega$ after taking the derivatives
in \eq\nr{rhoS_Psi}, 
so that the basis is orthogonal. Making use 
of \eqs\nr{nabla_rad}, \nr{Ylm},
the terms $m=\pm 1$ both yield
\ba
 \nabla_{\vec{r}'}\cdot \nabla_{\vec{r}}\; \delta_1 \tilde \Psi
 & = &  
 A \frac{3}{8\pi}
 \biggl\{
   \frac{\partial}{\partial r}
   \biggl[ 
     \frac{g_{<}^{1}(r)}{r}
   \biggr]   
   \frac{\partial}{\partial r'}
   \biggl[ 
     \frac{g_{>}^{1}(r')}{r'}
   \biggr] \sin^2\!\theta + 
     \frac{g_{<}^{1}(r)}{r^2}
     \frac{g_{>}^{1}(r')}{(r')^2}
     \Bigl[ \cos^2\!\theta + 1 \Bigr]
 \biggr\}
 \;, \hspace*{1cm}
\ea
while the term $m=0$ yields
\ba
 \nabla_{\vec{r}'}\cdot \nabla_{\vec{r}}\; \delta_1 \tilde \Psi
 & = & 
 A \frac{3}{4\pi}
 \biggl\{
   \frac{\partial}{\partial r}
   \biggl[ 
     \frac{g_{<}^{1}(r)}{r}
   \biggr]   
   \frac{\partial}{\partial r'}
   \biggl[ 
     \frac{g_{>}^{1}(r')}{r'}
   \biggr] \cos^2\!\theta + 
     \frac{g_{<}^{1}(r)}{r^2}
     \frac{g_{>}^{1}(r')}{(r')^2}
     \sin^2\!\theta 
 \biggr\}
 \;.
\ea
Summing together, we get
\ba
 \nabla_{\vec{r}'}\cdot \nabla_{\vec{r}} \; \delta_1 \tilde \Psi
 & = & 
 A \frac{3}{4\pi}
 \biggl\{
   \frac{\partial}{\partial r}
   \biggl[ 
     \frac{g_{<}^{1}(r)}{r}
   \biggr]   
   \frac{\partial}{\partial r'}
   \biggl[ 
     \frac{g_{>}^{1}(r')}{r'}
   \biggr]  + 
     2 \frac{g_{<}^{1}(r)}{r^2}
     \frac{g_{>}^{1}(r')}{(r')^2}
 \biggr\}
 \;.
\ea
We now insert 
$
 g^1_{<}(r) =  g^1_\rmi{r}(r)
$,
$
 g^1_{>}(r') = - g^1_\rmi{r}(r') 
 \int_{r'}^\infty \! {\rm d} r'' \, {1}/{[g^1_\rmi{r}(r'')]^2}
$
from \eqs\nr{glr2}, \nr{gli3}, and recall that at small $r$, 
$
 g^1_\rmi{r}(r) \approx \varrho^2 = (r  \alpha   M)^2 
$. % corresponding to the choice $a_0 = (\alpha M)^2$.
Thereby 
\be
 \lim_{\vec{r},\vec{r}'\to\vec{0}}  
 \nabla_{\vec{r}'}\cdot \nabla_{\vec{r}} \; \delta_1 \tilde \Psi
 = %% & = & 
 - \frac{3 A}{4\pi} (\alpha M)^3
 \lim_{\varrho'\to 0}
 \biggl\{
   \frac{{\rm d}}{{\rm d}\varrho'}
   \biggl(
    \frac{g_\rmi{r}^1(\varrho')}{\varrho'}
    \int_{\varrho'}^{\infty}
    \!  \frac{{\rm d}\varrho''}{[g_\rmi{r}^1(\varrho'')]^2} 
   \biggr)
   +  %% \nn  &&  \hspace*{3cm} +
    \frac{2 g_\rmi{r}^1(\varrho')}{(\varrho')^2}
    \int_{\varrho'}^{\infty}
    \!  \frac{{\rm d}\varrho'' }{[g_\rmi{r}^1(\varrho'')]^2} 
 \biggr\}
 \;.
\ee
Inserting this into \eq\nr{rhoS_Psi}, and 
making use of \eq\nr{AA}, we get 
\be
 \frac{\delta_1 \rho^S(\omega')}{M^2}  =  
 \frac{3\Nc}{2\pi} \alpha^3 
 \lim_{\varrho'\to 0}
 \im \biggl\{ 
 \biggl[
  \frac{{\rm d}}{{\rm d}\varrho'} 
  \biggl( \frac{g^1_\rmi{r}(\varrho')}{\varrho'} \biggr)
  + \frac{2 g^1_\rmi{r}(\varrho')}{(\varrho')^2}
 \biggr]
  \int_{\varrho'}^{\infty}
  \! 
  \frac{{\rm d}\varrho''}{[g^1_\rmi{r}(\varrho'')]^2}
 - \frac{1}{\varrho'  g^1_\rmi{r}(\varrho') }
  \biggr\} \;. \la{rhoP_1}
\ee 

Once again, we need to inspect the origin of the imaginary part. 
According to \eqs\nr{glr_expansion}, \nr{impart_expansion}, 
$
 \re [g^1_\rmi{r}(\varrho')] \sim (\rho')^2
$, 
$
 \im [g^1_\rmi{r}(\varrho')] \sim (\rho')^6
$, 
and consequently
\be
 \re \biggl\{
 \frac{1}{[g_\rmi{r}^1(\varrho'')]^2} 
 \biggr\} 
 \approx \frac{1}{(\varrho'')^4}
 \;, \quad
 \im \biggl\{
 \frac{1}{\varrho' g_\rmi{r}^1(\varrho') } 
 \biggr\}
 \approx  \im \biggl\{
 \frac{1}{(\varrho')^3 + i \gamma_1 (\varrho')^7 } 
 \biggr\}
 \approx
 - \gamma_1 \varrho'
 \;,  \la{ests_11}
\ee
so that the only possibility is to consider 
$
 \im\{ 
 {1} / {[g^1_\rmi{r}(\varrho'')]^2}
 \} 
$. 
The prefactor multiplying this can be trivially determined, 
and we end up with
\be
 \frac{\delta_1\rho^S(\omega')}{M^2}
= 
 \frac{9\Nc}{2\pi} \alpha^3 
 \left. \lim_{\delta\to 0}
 \int_{\delta}^{\infty}
  \! {\rm d}\varrho \,
  \im \left\{ \frac{1}{[g^1_\rmi{r}(\varrho)]^2} \right\}
 \right|_{g^1_\rmi{r}(\varrho) = \varrho^2 - \varrho^3/4 + ...}
 \;, \la{rhoP_2}
\ee
in analogy with \eq\nr{rho_res3}.
Combining \eqs\nr{new_3}, \nr{rhoP_2}, the complete
scalar channel spectral function can be written as in \eq\nr{rhoS_full}.

To conclude, let us again check that the procedure introduced does 
yield the correct tree-level result. Somewhat unfortunately, the 
first term in \eq\nr{rhoS_full} does not contribute in this limit: 
the subleading term in \eq\nr{glr_expansion} 
would be of $\rmO(\varrho^{l+3})$ in the free case, 
so that $g_\rmi{r}^0/\varrho\sim\varrho^2$ in \eq\nr{new_1}, 
and $\delta_0\rho^S$ vanishes. However, the second term
in \eq\nr{rhoS_full} survives. In dimensionless variables, 
the homogeneous Schr\"odinger equation reads
\be
 \biggl[ 
  \frac{{\rm d}^2}{{\rm d}\varrho^2}
 - \frac{2}{\varrho^2} + \frac{\hat\omega'}{\alpha^2} + i \epsilon
 \biggr] g^1_\rmi{r}(\varrho) = 0 
 \;. 
\ee
Since there is no imaginary potential, we have had to introduce 
$\epsilon \equiv 0^+$ to pick up the retarded solution. 
The solution normalised to give the desired small-$\varrho$ behaviour 
[$g^1_\rmi{r}(\varrho) = \varrho^2 + ...$] is
\be
 g^1_\rmi{r}(\varrho) =  \frac{3}{k^2}
 \biggl[ \frac{\sin (k\varrho) }{k \varrho} - \cos (k\varrho) \biggr]
 \;, 
\ee
where $k$ is from \eq\nr{k_def}.
We note that 
\be
% \frac{1}
 {
 \Bigl[ \frac{\sin (k\varrho)}{k \varrho} - \cos (k\varrho) \Bigr]^{-2}
 }
 = \frac{1}{k}  \frac{{\rm d}}{{\rm d}\varrho} \biggl[ 
 \frac{\cos (k\varrho) + k\varrho \sin (k\varrho) }
                  {k\varrho \cos (k\varrho) - \sin (k\varrho)}
 \biggr]
 \;,
\ee
whereby 
\ba
 \frac{\rho^S(\omega')}{M^2}
 & = &  
 \frac{9\Nc}{2\pi} \alpha^3 \lim_{\varrho\to\infty}
 \im\left\{
  \fr19 
  \biggl( \frac{\hat\omega'}{\alpha^2}+i \epsilon \biggr)^{\fr32}
  \frac{\cos ({k \varrho}) + {k \varrho} \sin ({k \varrho }) }
                  {{k \varrho} \cos ({k \varrho }) - \sin ({k \varrho })}
%   \frac{\cos\varrho\sqrt{\frac{\hat\omega'}{\alpha^2}+i \epsilon} + 
%   \varrho \sqrt{\frac{\hat\omega'}{\alpha^2}+i \epsilon} 
%   \sin \varrho \sqrt{\frac{\hat\omega'}{\alpha^2}+i \epsilon} }
%   {\varrho \sqrt{\frac{\hat\omega'}{\alpha^2}+i \epsilon} 
%    \cos \varrho \sqrt{\frac{\hat\omega'}{\alpha^2}+i \epsilon} - 
%   \sin \varrho \sqrt{\frac{\hat\omega'}{\alpha^2}+i \epsilon} } 
 \right\}
 \nn & = & 
  \frac{\Nc}{2\pi} \alpha^3 
  \im\biggl\{
  \biggl( \frac{\hat\omega'}{\alpha^2}+i \epsilon \biggr)^{\fr32}
  i 
  \biggr\} 
  = 
  \frac{\Nc}{2\pi} \; \theta(\omega') \, (\hat \omega')^{\fr32}
  \;.
\ea
This indeed agrees with \eq\nr{tree_P}.

%%%%%%%%%%%%%%%%%%%%%%%%%%%%%%%%%%%%%%%%%%%%%%%%%%%%%%%%%%%%%%%%%%%%%%%%%%%

\end{document}